\documentclass[prd,10pt,twocolumn,tightenlines,superscriptaddress,numerical,notitlepage,showpacs,amsmath,amssymb,amsfonts,aps,nofootinbib,longbibliography,floatfix]{revtex4-1}
\usepackage[utf8]{inputenc}
\usepackage[justification=raggedright,singlelinecheck=false,font=small]{caption}
\usepackage{graphicx,bm,bbm,color,amssymb,amsmath}
\usepackage{subcaption}
\usepackage{slashed,verbatim}
\usepackage{soul}
\usepackage[usenames,dvipsnames]{xcolor}
\sethlcolor{red}
\usepackage[colorlinks=true,linktocpage=true,linkcolor=blue,citecolor=blue]{hyperref}
\usepackage{ulem}
\usepackage{combelow}
\usepackage{dsfont}
\usepackage{amsmath,amssymb,mathtools,bm,bbm,braket,slashed}
\usepackage{tikz}
\usepackage{cancel}
\usepackage{xcolor}
\usepackage{ulem}
\usepackage{hyperref}
\hypersetup{colorlinks=true, citecolor=blue, urlcolor=blue, citecolor=red}

\definecolor{purple}{rgb}{0.8,0,0.6}

\definecolor{battleshipgrey}{rgb}{0.2, 0.52, 0.51}
\definecolor{darkgreen}{rgb}{0.12, 0.5, 0.17}

\renewcommand\sout{\bgroup\color{blue} \ULdepth=-.5ex \ULset}

    \newcommand{\beqn}{\begin{eqnarray}}
    \newcommand{\eeqn}{\end{eqnarray}}
    \newcommand{\beqs}{\begin{subequations}}
    \newcommand{\eeqs}{\end{subequations}\\[-2mm]\noindent}

    \newcommand{\nn}{\nonumber}

\begin{document}
\title{Dilepton Production as a Probe of Pion Condensation in Hot and Dense QCD Matter}

\author{Aritra Bandyopadhyay}
\email{aritra.bandyopadhyay@e-uvt.ro} 
\affiliation{Institute of Theoretical Physics, University of Wrocław, plac Maksa Borna 9, PL-50204 Wrocław, Poland}
\affiliation{Department of Physics, West University of Timişoara, Bd. Vasile Pârvan 4, Timişoara 300223, Romania}

\author{Chowdhury Aminul Islam}
\email{caislam.phys@aliah.ac.in}
\affiliation{Department of Physics, Aliah University, II-A/27, Action Area II, Newtown, Kolkata-700160, India.}

\author{Krzysztof Redlich}
\email{krzysztof.redlich@uwr.edu.pl}
\affiliation{Institute of Theoretical Physics, University of Wrocław, plac Maksa Borna 9, PL-50204 Wrocław, Poland}
\affiliation{Polish Academy of Sciences PAN, Podwale 75, PL-50449 Wroclaw, Poland}

\author{Chihiro Sasaki}
\email{chihiro.sasaki@uwr.edu.pl}
\affiliation{Institute of Theoretical Physics, University of Wrocław, plac Maksa Borna 9, PL-50204 Wrocław, Poland}
\affiliation{International Institute for Sustainability with Knotted Chiral Meta Matter (WPI-SKCM$^2$), Hiroshima University, 1-3-1 Kagamiyama, 739-8531, Higashi-Hiroshima, Hiroshima, Japan}

\date{\today}

\begin{abstract}
We investigate dilepton production from an isospin-asymmetric hot and dense medium in order to explore the role of isospin imbalance in electromagnetic spectral properties. We focus in particular on modifications of the dilepton production rate associated with the onset of pion condensation, which can occur in the presence of a finite isospin chemical potential. We employ the Nambu--Jona-Lasinio model with isoscalar--vector interaction. We examine the phase structure in the $T-\mu_I$ plane and estimate the vector current correlator--resummed dilepton rate for an effective quark chemical potential. We find that the interplay between isospin asymmetry, pion condensation, and vector interactions leads to nontrivial modifications of the dilepton yield. In particular, we observe two key features of the pion condensed phase: an enhancement at lower invariant mass and a prominent plateau-like structure which also help clearly identify the pion condensed phase from a chirally broken/restored phase. These results highlight the potential sensitivity of dilepton observables to pion-condensed phase of QCD matter, with possible implications for future low-energy heavy-ion collision experiments as well as isospin-rich environments such as neutron star matter.
\end{abstract}

\maketitle

\section{Introduction}

Dilepton emission has long been recognized as a key electromagnetic probe of strongly interacting matter produced in relativistic heavy-ion collisions~\cite{Shuryak:1978ij,STAR:2013pwb,PHENIX:2015vek}. Due to their negligible final-state interactions, dileptons escape the medium largely unaltered, encoding information about the entire space--time evolution of the fireball. Since their production rate is directly proportional to the imaginary part of the in-medium vector current--current correlator~\cite{McLerran:1984ay,Kajantie:1986dh,Weldon:1990iw}, dileptons provide a sensitive window into the spectral properties of quarks and hadrons under extreme conditions~\cite{Rapp:1999ej,Rapp:2014hha,NA60:2006ymb}.

In realistic collision environments, particularly those involving neutron-rich nuclei or lower beam energies, the produced medium can exhibit significant isospin asymmetry. This asymmetry is naturally characterized by a finite isospin chemical potential, $\mu_I$, which leads to unequal number densities of up and down quarks~\cite{Son:2000xc}. Alongside the temperature $T$ and baryon chemical potential $\mu_B$, a nonzero $\mu_I$ influences the self-consistent dynamical mass generation in chiral effective models. Consequently, the effective quark mass and the associated in-medium spectral functions develop a nontrivial dependence on the thermodynamic parameters. These modifications arising from finite isospin asymmetry are also reflected in the structure of the QCD phase diagram, particularly at high baryon density~\cite{Fukushima:2013rx,Aarts:2023vsf}. Moreover, lattice QCD studies at finite isospin density are free from the notorious sign problem~\cite{Kogut:2001id,Kogut:2002zg,Brandt:2017oyy,Brandt:2018bwq}, providing a reliable benchmark for exploring the $T$–$\mu_I$ plane of the QCD phase diagram.

The Nambu--Jona-Lasinio (NJL) model~\cite{Nambu:1961tp,Nambu:1961fr} provides a well-established theoretical framework for investigating these effects. Owing to its chiral symmetry structure and dynamical mass generation mechanism, the model allows a transparent study of how the quark propagator, mesonic modes, and vector current correlators respond to changes in $T$, $\mu_B$, and $\mu_I$. 
In particular, the NJL model naturally exhibits the critical behavior associated with pion condensation: when $\mu_I$ exceeds the in-medium pion mass, the system undergoes a phase transition into a state characterized by a nonzero charged-pion condensate~\cite{Turko:1993dy,Abuki:2008wm,Andersen:2007qv,Sun:2007fc,Ebert:2005wr,Ebert:2005cs,
He:2006tn,He:2005nk,He:2005sp,Barducci:2004tt,Toublan:2003tt,
Frank:2003ve,Mu:2010zz,Xia:2013caa}. See also Refs.~\cite{Ebert:2016hkd,Khunjua:2017khh,
Khunjua:2019lbv,Khunjua:2019ini,Khunjua:2020xws,Lu:2019diy,Avancini:2019ego,Lopes:2021tro} for more recent developments on the same. 

In addition to scalar and pseudoscalar interaction channels, vector interactions can also play an important role in determining the properties of dense QCD matter~\cite{Klevansky:1992qe,Buballa:2003qv,Fukushima:2008wg,Kojo:2014rca,Ali:2024owl}, including in the presence of a finite isospin chemical potential $\mu_I$~\cite{Sasaki:2006ws}. The inclusion of an isoscalar–vector four-fermion interaction in the NJL model leads to a density-dependent shift of the effective quark chemical potential. At the same time, it modifies the vector current correlator through the resummation of vector interaction channels~\cite{Davidson:1995fq,Islam:2014sea}. These effects are expected to impact both the phase structure and the electromagnetic spectral function, thereby influencing the dilepton production rate (DPR) in a nontrivial manner~\cite{Rapp:1999ej,Rapp:2014hha,Islam:2014sea}.

Therefore, the onset of a pion-condensed phase is expected to leave a distinctive imprint on the dilepton yield~\cite{He:2005nk,Abuki:2008wm,Rapp:1999ej}. Modifications to the effective quark mass and altered dispersion relations of pseudoscalar and vector mesons affect the electromagnetic spectral function. The presence of vector interactions further modifies these features through changes in the effective quark chemical potential and the structure of the vector current correlator~\cite{Davidson:1995fq,Sasaki:2006ws,Islam:2014sea}. Consequently, the DPR emerges as a promising observable for identifying the presence and onset of pion condensation. Within the NJL framework, these effects can be systematically incorporated through the dynamically generated effective quark mass and chemical potential~\cite{Klevansky:1992qe,Buballa:2003qv}.

Thus, studying dilepton emission in an isospin-asymmetric medium within the NJL model provides not only a theoretically controlled way to quantify the impact of $\mu_I$ on electromagnetic radiation, but also a concrete pathway to establishing dileptons as a potential signature of pion condensation. This investigation becomes particularly relevant for ongoing and future experimental programs, such as FAIR~\cite{CBM:2016kpk}, J-PARC~\cite{Sako:2014fha} and NICA~\cite{Kekelidze:2017tgp}, where strong isospin imbalance and moderate temperatures may create favorable conditions for exploring this intriguing region of the QCD phase diagram. 

The paper is organized as follows. In Sec.~\ref{sec:formalism_dpr}, we present the theoretical framework employed in this work. This section is further divided into two subsections, where we first introduce the NJL model and then discuss the formulation of the DPR including the resummed vector current correlator. In Sec.~\ref{sec:results}, we present our numerical results and discuss the corresponding physical implications. Finally, we summarize our findings and present our conclusions in Sec.~\ref{sec:conclusion}.

\section{Formalism}
\label{sec:formalism_dpr}
In order to consistently incorporate medium effects arising from finite temperature $(T)$, baryon chemical potential $(\mu_B)$, and isospin asymmetry into the DPR, we employ the two-flavor Nambu--Jona-Lasinio (NJL) model as an effective description of strongly interacting matter in the nonperturbative regime. The NJL framework dynamically generates a constituent quark mass through chiral symmetry breaking and also allows for the possibility of pion condensation at relatively higher values of finite isospin chemical potential ($\mu_I$). 

In addition to the usual scalar and pseudoscalar interactions, we include an isoscalar-vector interaction channel. In a dense medium, the inclusion of the vector interaction is crucial, as it couples with the number density and effectively shifts the quark chemical potential, which is dictated by the strength of the vector interaction~\cite{Fukushima:2008wg,Kojo:2014rca,Ali:2024owl}.

In the present context, the medium modifications relevant for dilepton production in an isospin asymmetric hot and dense medium therefore arise through:
(i) the dynamically generated quark mass, which depends explicitly on the chiral condensate and implicitly on the pion condensate,
(ii) the shift of the effective chemical potential induced by the isoscalar-vector interaction,
(iii) the presence of finite isospin asymmetry, and
(iv) the modification of the in-medium vector current correlator through vector-channel resummation obtained through random phase approximation (RPA).

In the following, we first present the formulation of the NJL model at finite $T,\mu_B \;\rm{and}\;\mu_I$ involving scalar, pseudoscalar and isoscalar-vector channels and then describe how the resulting medium-modified quasiparticle properties are implemented in the vector-channel resummed DPR.


\subsection{The SU(2) NJL model at finite isospin chemical potential}
\label{subsec:su2_njl}

\begin{figure*}[t]
    \centering
    \includegraphics[scale=0.8]{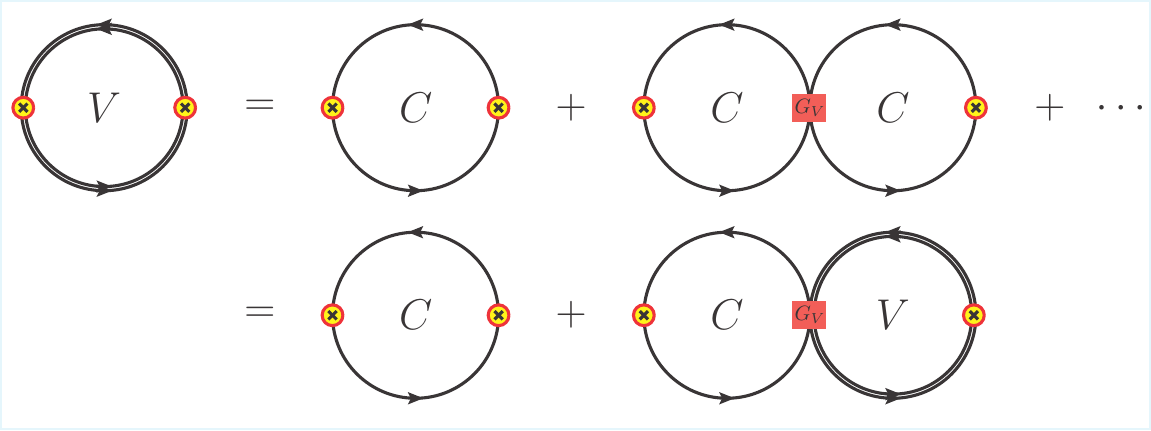}
    \caption{Diagrammatic representation of the RPA (ring) resummation for the vector current correlator in the presence of an isoscalar-vector interaction $G_V$. The circles labeled $C$ and double-lined circles labeled $V$ denote the one-loop (bare) and fully resummed vector current correlators, respectively. The circled crosses ($\otimes$) represent the external vector current insertion points of the correlator. The squared $G_V$ vertex ($\boxed{G_V}$) represents the isoscalar-vector interaction entering the resummation. The first line shows the explicit geometric series expansion $V = C + C\,G_V\,C + \cdots$, while the second line represents the compact Dyson--Schwinger form $V = C + G_V\,C\,V$, corresponding to Eq.~\eqref{eq:V_munu} of the text. 
    }
    \label{fig:vector_resummed}
\end{figure*}

We consider a two-flavor Nambu--Jona-Lasinio (NJL) model at finite temperature, baryon chemical potential, and isospin chemical potential. The Lagrangian density that we choose to work with contains scalar, pseudoscalar, and isoscalar-vector four-fermion interactions
\begin{align}
\mathcal{L}_{\mathrm{NJL}} &= \bar{\psi}(i\slashed{\partial}-m+\hat{\mu}\,\gamma_0)\psi + G_S\left[(\bar{\psi}\psi)^2 + (\bar{\psi}i\gamma_5\vec{\tau}\psi)^2\right] \nonumber\\
&- G_V(\bar{\psi}\gamma_\mu\psi)^2, \nonumber\\
&= \bar{\psi}(i\slashed{\partial}-m+\hat{\mu}\,\gamma_0)\psi+ G_S\Big[(\bar{\psi}\psi)^2
+ (\bar{\psi}i\gamma_5\tau_3\psi)^2\nonumber \\
&+ 2(\bar{\psi}i\gamma_5\tau_+\psi)(\bar{\psi}i\gamma_5\tau_-\psi)\Big] - G_V(\bar{\psi}\gamma_\mu\psi)^2,
\label{eq:lag_njl}
\end{align}
where $\psi$ denotes the quark doublet containing the light quarks, $m$ is the current quark mass, and $G_S$ and $G_V$ are the scalar and isoscalar-vector coupling constants, respectively. The matrices $\tau_i$ are the Pauli matrices in flavor space, and $\tau_\pm=(\tau_1\pm i\tau_2)/\sqrt{2}$. The chemical potential matrix in flavor space is given by
\begin{equation}
\hat{\mu} =
\begin{pmatrix}
\mu_u & 0 \\
0 & \mu_d
\end{pmatrix},
\end{equation}
with the light-quark chemical potentials defined as
\begin{equation}
\mu_u = \frac{\mu_B}{3}+\frac{\mu_I}{2}\qquad {\rm and} 
\qquad
\mu_d = \frac{\mu_B}{3}-\frac{\mu_I}{2}.
\end{equation}

A finite isospin chemical potential explicitly breaks the $SU(2)$ isospin symmetry down to $U(1)_{I_3}$. Within the mean-field approximation, we impose the ansatz
\begin{equation}
\langle \bar{\psi} i\gamma_5 \tau_3 \psi \rangle = 0,
\end{equation}
thereby allowing for charged pion condensation.

We define the chiral and charged pion condensates as
\begin{equation}
\sigma = -2G_S\,\langle\bar{\psi}\psi\rangle,
\end{equation}
\begin{align}
\sqrt{2}\,\pi_+ &= -2\sqrt{2}\,G_S\,
\langle\bar{\psi}i\gamma_5\tau_+\psi\rangle
= \Delta\,e^{i\theta}, \\
\sqrt{2}\,\pi_- &= -2\sqrt{2}\,G_S\,
\langle\bar{\psi}i\gamma_5\tau_-\psi\rangle
= \Delta\,e^{-i\theta},
\end{align}
where $\theta$ characterizes the spontaneous breaking of $U(1)_{I_3}$.

The isoscalar-vector mean field is defined as
\begin{equation}
\Sigma_V = 2G_V n,
\qquad
n=\langle\bar{\psi}\gamma_0\psi\rangle,
\end{equation}
with $n$ denoting the quark number density.

The mean-field thermodynamic potential is then given by
\begin{multline}
\Omega_{\mathrm{NJL}}(\sigma,\Delta,\Sigma_V,T)
= \frac{\sigma^2+\Delta^2}{4G_S}
-\frac{\Sigma_V^2}{4G_V}
- 2N_c \int\limits_k^{\Lambda}\sum_{s=\pm} E_k^s \\
- 2N_c\int\limits_k\sum_{s=\pm}
T\ln\!\left[\left(1+e^{-\beta(E_k^s+\tilde{\mu})}\right)
\left(1+e^{-\beta(E_k^s-\tilde{\mu})}\right)\right],
\label{Omega_NJL}
\end{multline}
where
\begin{equation}
E_k^\pm = \sqrt{(E_k \pm \mu_I/2)^2 + \Delta^2},
\qquad
E_k = \sqrt{k^2 + M^2},
\end{equation}
with $M = m + \sigma$ and the effective chemical potential
\begin{equation}
\tilde{\mu} = \frac{\mu_B}{3}-\Sigma_V.
\end{equation}
The notation $\int_k^\Lambda$ denotes three-momentum integration with a sharp cutoff $\Lambda$. Note that $\tilde{\mu}$ here carries only the baryonic shift due to the vector mean field, while the isospin splitting is absorbed into the Nambu--Gorkov quasi-particle energies $E_k^\pm$; this is to be contrasted with the flavor-dependent effective chemical potential $\tilde{\mu}_f$ introduced in subsection~\ref{subsec:dpr}.

The equilibrium values of the condensates (denoted by subscript $m$) are obtained from the gap equations :
\begin{equation}
\left.
\frac{\partial \Omega_{\mathrm{NJL}}}{\partial \sigma}
\right|_{\sigma=\sigma_m}
=
\left.
\frac{\partial \Omega_{\mathrm{NJL}}}{\partial \Delta}
\right|_{\Delta=\Delta_m}
=
\left.
\frac{\partial \Omega_{\mathrm{NJL}}}{\partial \Sigma_V}
\right|_{\Sigma_V=\Sigma_{V,m}}
= 0.
\end{equation}


\subsection{DPR in an isospin asymmetric dense medium}
\label{subsec:dpr}

In this work, we assume that the system created in a heavy-ion collision is already in local thermal equilibrium. We focus on the production of on-shell lepton pairs from a generic process of the form
\begin{equation}
I \to l\,\bar{l} + F ,
\end{equation}
where $I$ and $F$ denote the initial and final asymptotic quark--gluon states, respectively.

For such a thermalized medium with $N_f=2$ flavors, the finite-temperature differential DPR for a virtual photon with four-momentum $Q^\mu = (\omega,\vec{q})$ can be written as~\cite{Karsch:2000gi}
\begin{equation}
\frac{dR}{d^4x\,d^4Q} = \frac{5 \alpha^2}{27 \pi^2} \frac{1}{Q^2} \, 
\frac{1}{e^{\omega/T} - 1} \, \rho_V(\omega, \vec{q}) \,,
\label{eq:dr_expr}
\end{equation}
where $Q^2=\omega^2-q^2$ is the invariant mass squared of the virtual photon, $\alpha$ is the fine-structure constant, and $\rho_V$ is the vector spectral function, which is related to the imaginary part of the vector current correlator. It governs not only the dilepton rate but also important transport and screening properties of the medium, such as the electrical conductivity and the Debye screening mass. 

However, the above equation is strictly valid for massless lepton pairs and in an isospin symmetric environment. If that symmetry is lost, whether by a background magnetic field or by an isospin chemical potential, as in the case in hand, the flavor sum can no longer be carried out trivially due to the unequal treatment of the $u$ and $d$ quarks~\cite{Mustafa:2025uad}. Thus, for nonzero $\mu_I$, Eq.~(\ref{eq:dr_expr}) takes the form
\begin{align}
    \frac{dR}{d^4x\,d^4Q} = \frac{\alpha}{12 \pi^3} \frac{1}{Q^2} \, 
    \frac{1}{e^{\omega/T} - 1} \, \sum\limits_{f=u,d}q_f^2\,\rho_{V;f}(\omega, \vec{q}) \,,
    \label{eq:dr_expr_nonzeromuI}
\end{align}
where $q_f$ and $\rho_{V;f}$ are the electric charge and single-flavor vector spectral function, respectively, corresponding to a particular quark flavor $f$. In the isospin symmetric limit $\mu_I \to 0$, one recovers $\rho_{V;u} = \rho_{V;d} \equiv \rho_V$ and $\sum_{f=u,d} q_f^2 = 5/9\,e^2$, with $e^2=4\pi\alpha$, so that 
Eq.~\eqref{eq:dr_expr_nonzeromuI} consistently reduces to Eq.~\eqref{eq:dr_expr}.

The isoscalar-vector interaction $G_V$ plays a particularly significant role in a dense matter, as it couples directly to the quark number density, $n = \langle\bar{\psi}\gamma^0\psi\rangle$, thereby modifying the in-medium vector current correlator~\cite{Islam:2014sea}. This interaction affects chiral symmetry restoration, the location of the critical endpoint, and the equation of state. Since the DPR is directly proportional to the vector spectral function, consideration of a finite $G_V$ requires a consistent resummation of the vector correlator to account for its direct effect. 

Such a resummation can be carried out by summing the geometric series 
of one-loop irreducible diagrams~\cite{Davidson:1995fq, Islam:2014sea, 
Hatsuda:1994pi, Klevansky:1992qe, Kapusta:2006pm,Rapp:1999ej} (ring or random phase 
approximation (RPA)), as illustrated in Fig.~\ref{fig:vector_resummed}. In the present framework, the single-flavor Dyson--Schwinger equation for the full vector correlator reads
\begin{equation}
V_{\mu\nu ;f} = C_{\mu\nu ;f} + 2G_V C_{\mu\sigma; f} V^\sigma_{\ \nu ;f},
\label{eq:V_munu}
\end{equation}
where \(C_{\mu\nu;f}\) is the single-flavor one-loop vector correlator which can be written as
\begin{equation}
C_{\mu\nu;f}(Q)= \int \frac{d^4P}{(2\pi)^4} \textrm{Tr}_{D,c} 
\left[\gamma_\mu S_f(P+Q) \gamma_\nu S_f(P)\right],
\label{eq:oneloop_Cmunu}
\end{equation}
where $\textrm{Tr}_{D,c}$ denotes traces over Dirac and color indices 
for a single quark flavor $f$, and $S_f(P)$ is the quark propagator in 
the chosen effective model. For our purpose, we will use the NJL model (Hartree approximation) version of the same~\cite{Klevansky:1992qe,Hatsuda:1994pi,Buballa:2003qv}, i.e.
\begin{equation}
S_f(P) = \frac{1}{\slashed{P}-M+\gamma_0 \tilde \mu_f},
\end{equation}
where $M$ is the flavor-independent constituent quark mass and $\tilde \mu_f = \mu_B/3 - \Sigma_V 
+ I_f\,\mu_I/2$ (with $I_f = +1$ for $u$ and $I_f = -1$ for $d$) is the 
flavor-dependent effective chemical potential.

At finite temperature and chemical potential, the correlators decompose into transverse and longitudinal components~\cite{Kapusta:2006pm,Bellac:2011kqa} :
\begin{align}
\Pi_{\mu\nu} &= \Pi_T P^T_{\mu\nu} + \Pi_L P^L_{\mu\nu},
\end{align}
with projection operators
\begin{equation}
P^L_{\mu\nu} = \frac{Q^2}{\tilde Q^2} \bar u_\mu \bar u_\nu, \quad
P^T_{\mu\nu} = \eta_{\mu\nu} - u_\mu u_\nu - \frac{\tilde Q_\mu \tilde Q_\nu}{\tilde Q^2}.
\end{equation}
Here, $\Pi$ represents both $C$ (one-loop) and $V$ (resummed) correlators, the heat bath velocity $u_\mu$ is $(1,0,0,0)$ in the rest frame of the medium, $\tilde Q_\mu = Q_\mu - \omega u_\mu$, and \(\bar u_\mu = u_\mu - \omega Q_\mu/Q^2\). For $D=4$, the corresponding coefficients relate to the temporal and spatial components as follows, where $\Pi_{ii} \equiv \Pi_{11}+\Pi_{22}+\Pi_{33}$ denotes the spatial trace of the correlator:
\begin{align}
    \Pi_L = - \frac{Q^2}{q^2}\Pi_{00}, \quad \Pi_T = \frac{1}{2}\left[\frac{\omega^2}{q^2}\Pi_{00}-\Pi_{ii}\right].
    \label{eq:coeff_to_compo}
\end{align}

Using Eq.~\eqref{eq:coeff_to_compo} and the fact that the scalar Dyson--Schwinger equation~\eqref{eq:V_munu} separates for transverse and longitudinal modes, i.e.
\begin{align}
V_T = \frac{C_T}{1 - 2G_V C_T}, \quad V_L = \frac{C_L}{1 - 2G_V C_L},
\end{align}
respectively, the temporal and spatial components of the contracted resummed correlator $V_{\mu\mu}$ can be expressed as:
\begin{align}
V_{00} &= \frac{C_{00}}{1 + 2G_V \frac{Q^2}{q^2} C_{00}}, \\
 V_{ii}&= \frac{C_{ii}-\frac{\omega^2}{q^2}C_{00}}{1-
 G_V(\frac{\omega^2}{ q^2}C_{00}-C_{ii})}
 + \frac{\frac{\omega^2}{ q^2}C_{00}}
 {1+2G_V\frac{Q^2}{ q^2}C_{00}}.
\end{align}

Corresponding imaginary parts along with the real and imaginary parts of the 
single-flavor one-loop vector current-current correlator in the presence of a 
finite isospin chemical potential are given in detail in appendix~\ref{appA}. 
Finally, the single-flavor resummed vector spectral function reads
\begin{equation}
\rho_{V;f} = \frac{1}{\pi} \left[ V^{\rm I}_{00;f} - V^{\rm I}_{ii;f} \right],
\label{eq.spectral_resum}
\end{equation}
where $V^{\rm I}_{00;f}$ and $V^{\rm I}_{ii;f}$ are imaginary parts of the resummed vector current correlator and computed using the single-flavor correlator expressions detailed in appendix~\ref{appA}. 
The full DPR in a hot and dense matter is then obtained by inserting $\rho_V^f$ 
into Eq.~\eqref{eq:dr_expr_nonzeromuI} and summing over flavors weighted 
by $q_f^2$. In the following section, we present numerical results for the DPR obtained using the NJL-dressed quark mass and analyze the 
impact of finite isospin asymmetry and pion condensation on the dilepton spectra.

\begin{figure*}[t]
    \centering
    \includegraphics[scale=0.4]{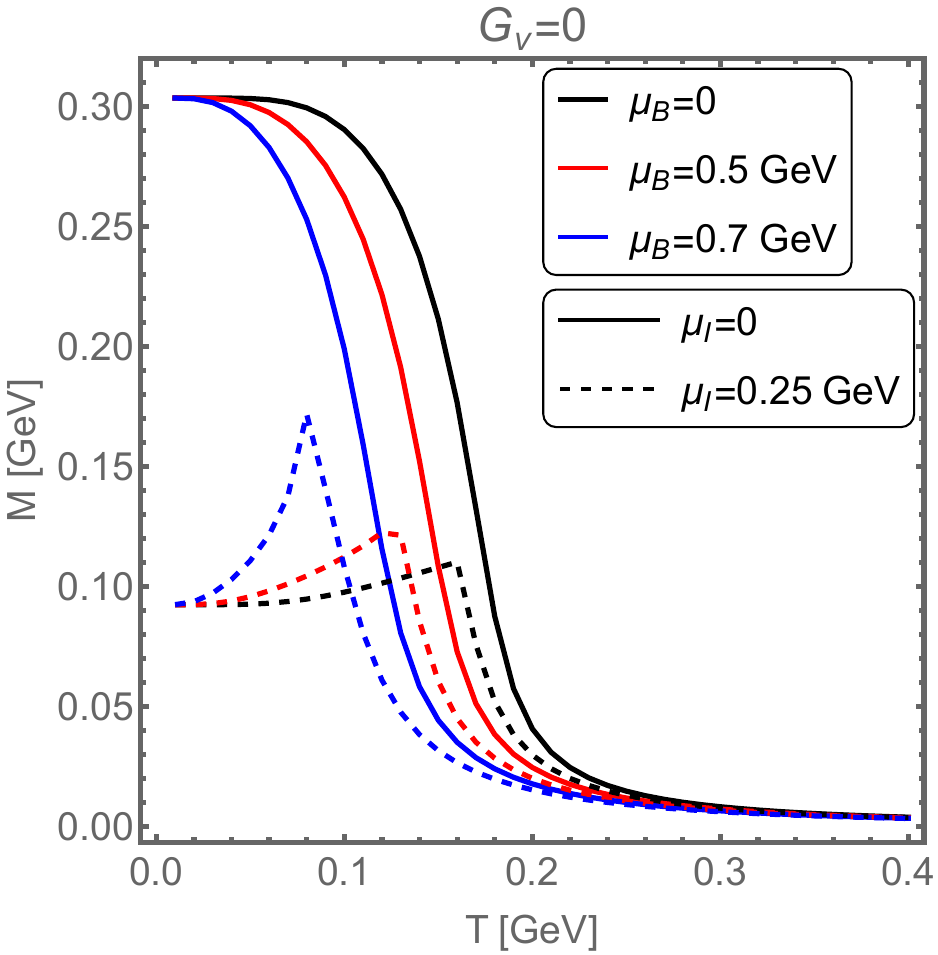}\hspace{1cm}
    \includegraphics[scale=0.4]{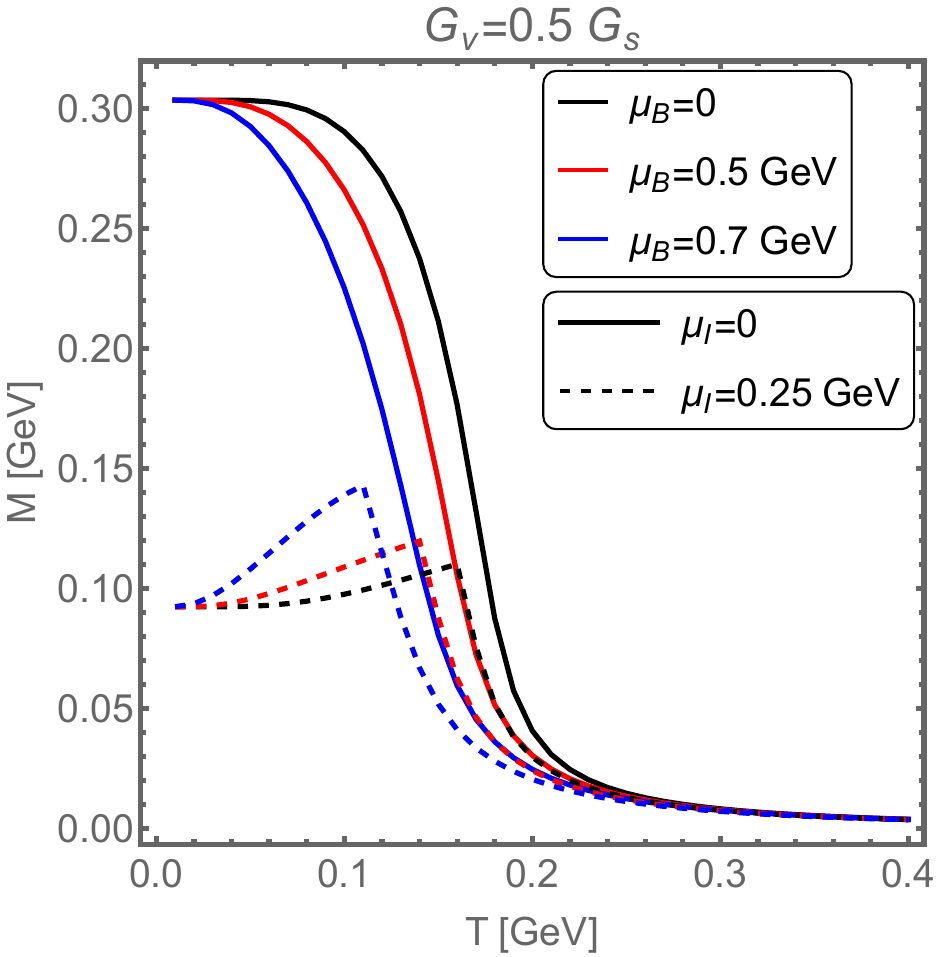}\\
    \vspace{0.5cm}
    \includegraphics[scale=0.4]{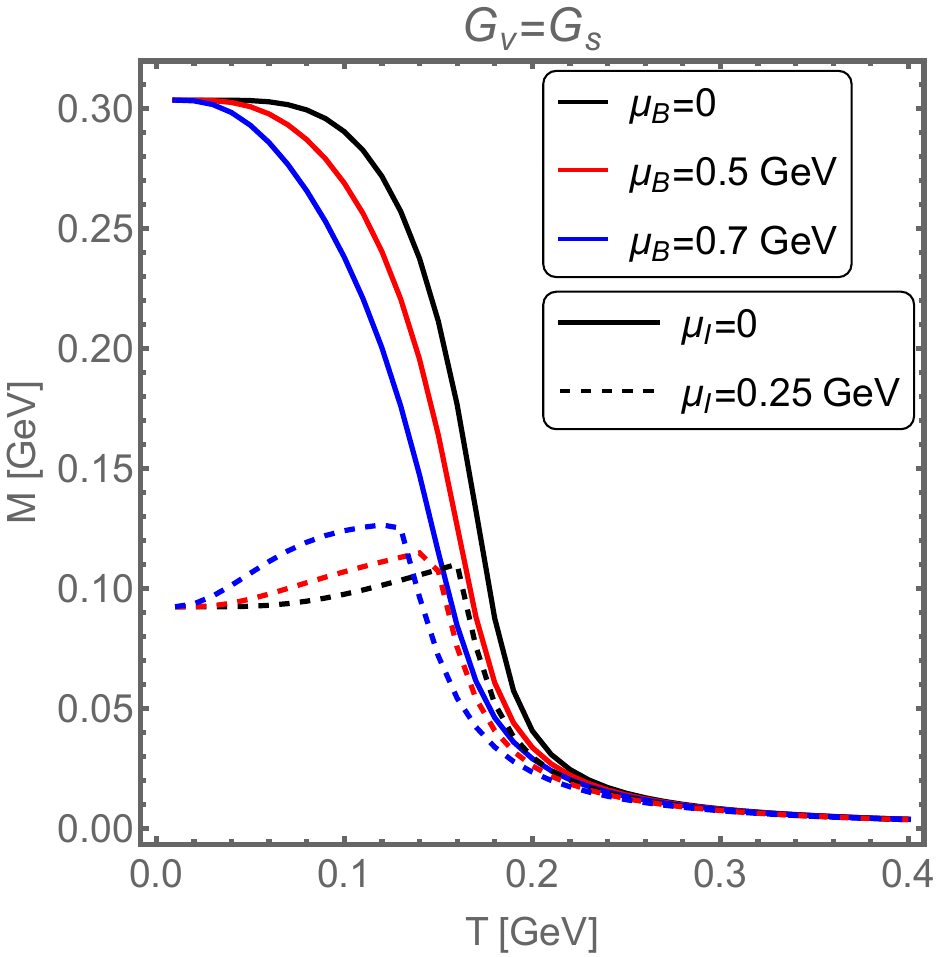}\hspace{1cm}
    \includegraphics[scale=0.4]{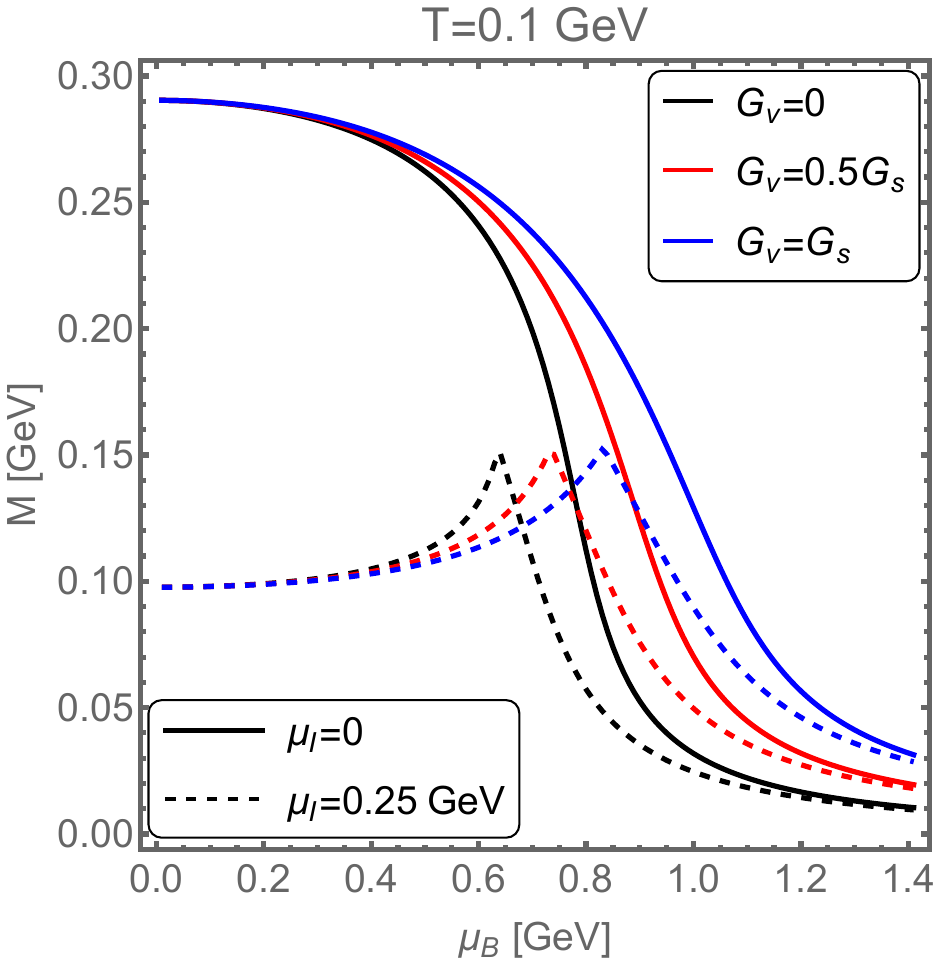}
    \caption{The effective quark mass $M$, a key input to the DPR, as a function of temperature $T$ (upper-left, upper-right, and lower-left panels) and baryon chemical potential $\mu_B$ (lower-right panel) within the SU(2) NJL model at finite isospin asymmetry, for three values of the isoscalar-vector coupling: $G_V = 0$, $0.5\,G_S$, and $G_S$. In the first three panels, black, red, and blue curves correspond to $\mu_B = 0$, $0.5$~GeV, and $0.7$~GeV, respectively, while in the lower-right panel the same color coding corresponds to the three $G_V$ values in increasing order. In all panels, solid and dashed lines correspond to $\mu_I = 0$ and $\mu_I = 0.25$~GeV.
    }
    \label{fig:Mq_muImuBGv}
\end{figure*}

\section{Results}
\label{sec:results}

\begin{figure*}[t]
    \centering
    \includegraphics[scale=0.4]{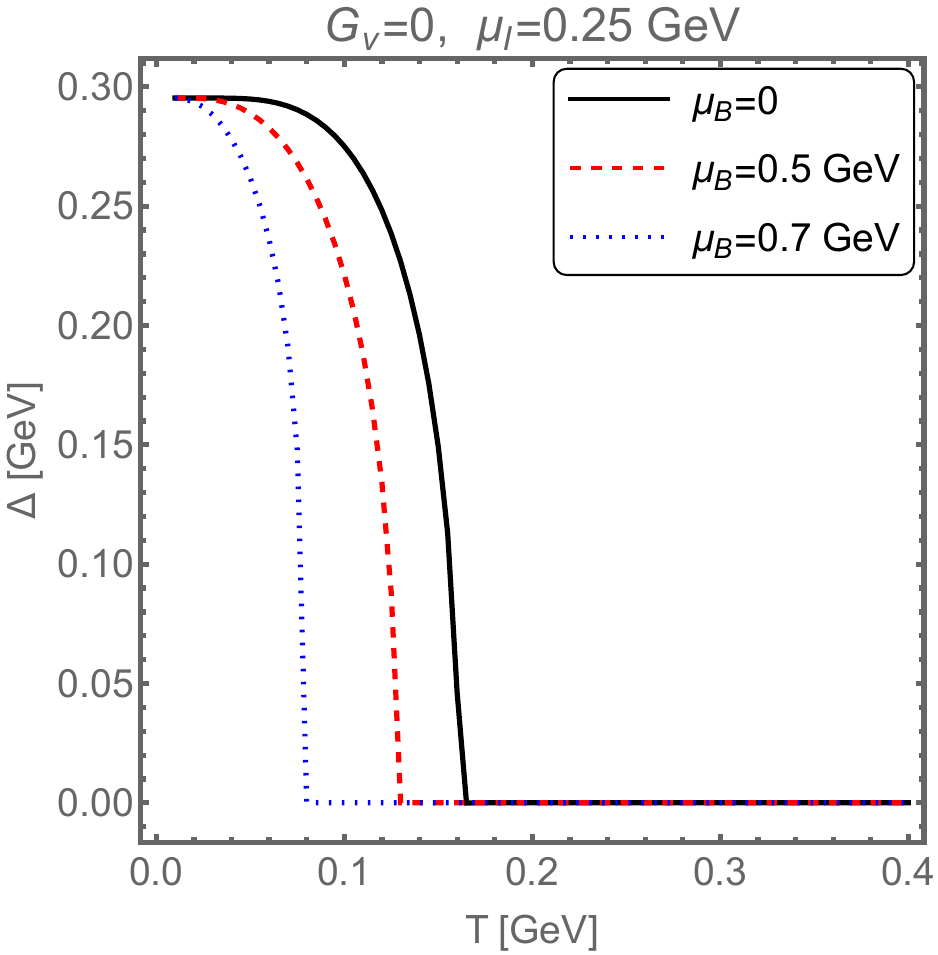}\hspace{1cm}
    \includegraphics[scale=0.4]{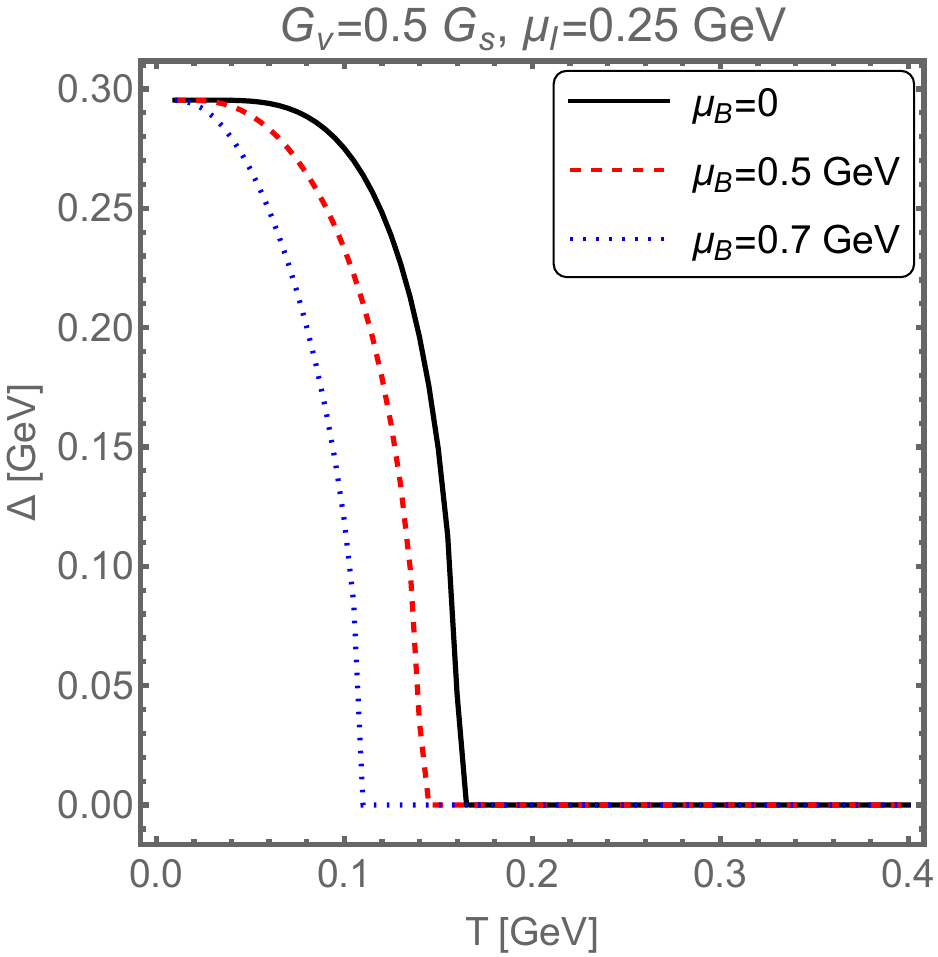}\\
    \vspace{0.5cm}
    \includegraphics[scale=0.4]{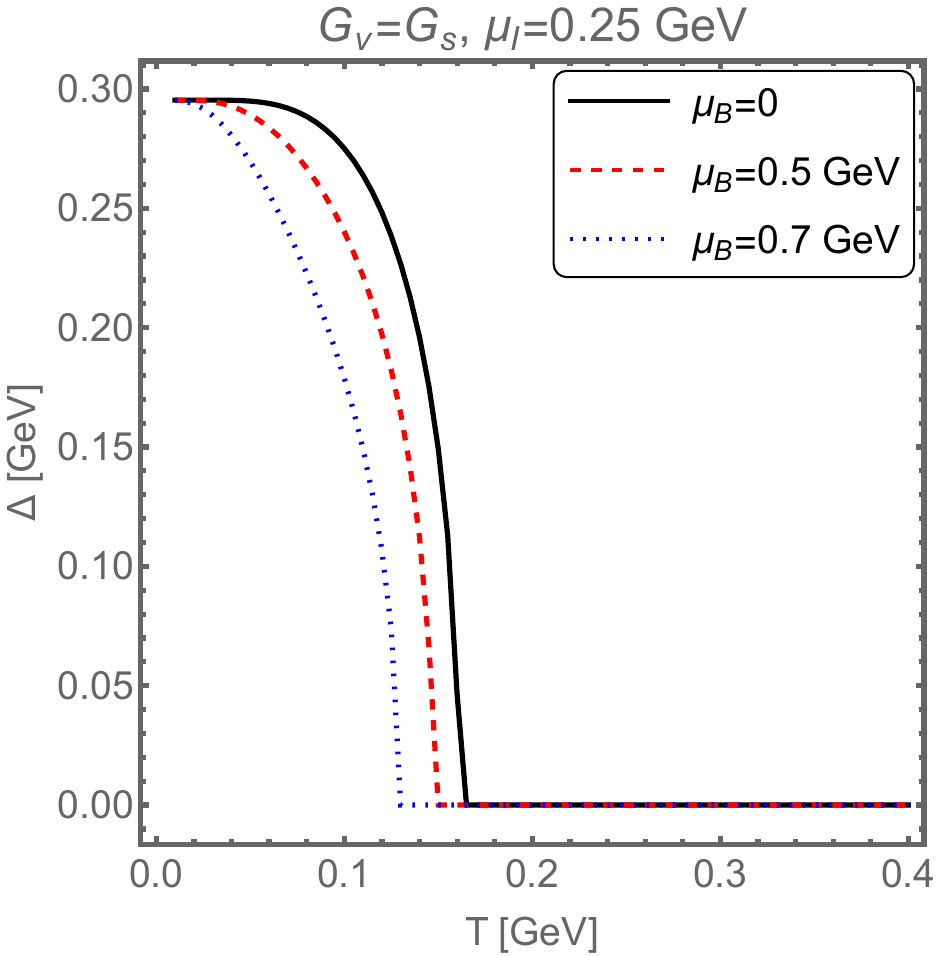}\hspace{1cm}
    \includegraphics[scale=0.4]{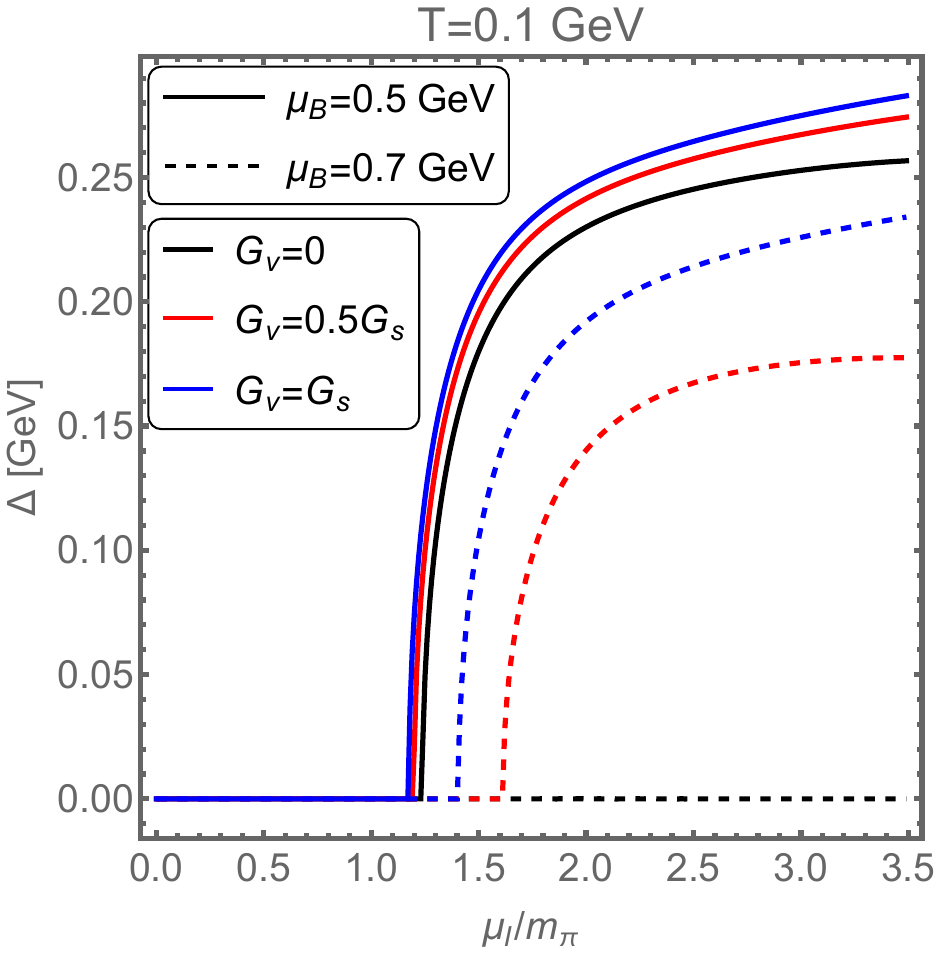}
    \caption{The pion condensate $\Delta$ as a function of temperature $T$ (upper-left, upper-right, and lower-left panels) and isospin chemical potential $\mu_I/m_\pi$ (lower-right panel) within the SU(2) NJL model, for $G_V = 0$ (upper-left), $G_V = 0.5\,G_S$ (upper-right), and $G_V = G_S$ (lower-left). In the first three panels, solid, dashed, and dotted curves correspond to $\mu_B = 0$, $0.5$~GeV, and $0.7$~GeV, respectively, at fixed $\mu_I = 0.25$~GeV. In the lower-right panel, the black, red, and blue curves correspond to the three $G_V$ values in increasing order, while solid and dashed lines correspond to $\mu_B = 0.5$~GeV and $\mu_B = 0.7$~GeV, respectively, at fixed $T = 0.1$~GeV.}
    \label{fig:Delta_muImuBGv}
\end{figure*}

The parameter set employed in the present study is $m = 4.76~\mathrm{MeV}$, $\Lambda = 659~\mathrm{MeV}$, and $G_S = 4.78~\mathrm{GeV}^{-2}$. These parameters are determined by fitting the pion mass value used in Lattice QCD~\cite{Brandt:2017oyy,Brandt:2018bwq}, namely $m_\pi = 131.7~\mathrm{MeV}$, together with $f_\pi = 92.4~\mathrm{MeV}$ and $\langle \bar{\psi}\psi \rangle^{1/3} = -250~\mathrm{MeV}$. This parameter set yields a vacuum constituent quark mass $M_0 \simeq 303.5~\mathrm{MeV}$~\cite{Klevansky:1992qe,Hatsuda:1994pi,Buballa:2003qv}. 

The remainder of this section is organized into three parts. First, we analyze the behavior of the order parameters (condensates) in an isospin-asymmetric hot and dense medium, including the effects of isoscalar-vector interactions. We then illustrate how these features are reflected in the $T$--$\mu_I$ plane of the QCD phase diagram for different values of the baryon chemical potential $\mu_B$ and the vector coupling $G_V$. Finally, we examine the cumulative impact of these effects on the DPR and attempt to identify possible signatures of pion condensation.

\subsection{Order Parameters}

Figure~\ref{fig:Mq_muImuBGv} shows the behavior of the effective quark mass $M$ as a function of $T$ and $\mu_B$ within the SU(2) NJL model. The two values of $\mu_I$ are chosen to contrast the isospin-symmetric $(\mu_I = 0)$ and pion-condensed $(\mu_I = 0.25~\mathrm{GeV})$ phases, with the pion condensation threshold being $\mu_I = m_\pi$ for $T=0$ (Silver-Blaze property~\cite{Cohen:2003kd}).

At vanishing isospin chemical potential, the effective mass exhibits the expected chiral crossover behavior~\cite{Aoki:2006we, Bhattacharya:2014ara, HotQCD:2014kol}, decreasing rapidly around a pseudo-critical temperature that shifts to lower values as $\mu_B$ increases. A finite $\mu_I$ modifies the temperature dependence nontrivially: it induces a non-monotonic structure in $M$, leading to a partial enhancement of the mass in the transition region before chiral restoration, which increases with $\mu_B$. This non-monotonicity arises from the competition between the suppression of the chiral condensate and the onset of the pion condensate (see Figure.~\ref{fig:Delta_muImuBGv} below), which partially compensates the mass reduction due to chiral transition. The magnitude and sharpness of this structure depend sensitively on the vector coupling strength.

The lower-right panel of Figure.~\ref{fig:Mq_muImuBGv} shows that increasing $G_V$ shifts the chiral transition to larger $\mu_B$, reflecting the stiffening effect of the vector interaction~\cite{Fukushima:2008wg,Bratovic:2012sd,Steinheimer:2010sp,Masuda:2015jka}. A finite $\mu_I$ again modifies the transition behavior, implying a visible deformation of the mass curve in the vicinity of the chiral crossover. Figure~\ref{fig:Mq_muImuBGv} overall illustrates that isospin asymmetry and vector interactions interplay nontrivially in determining the location and structure of the chiral transition within the SU(2) NJL framework, thereby significantly affecting the values of the effective quark mass, especially at lower temperatures.

\begin{figure*}[t]
    \centering
    \includegraphics[scale=0.35]{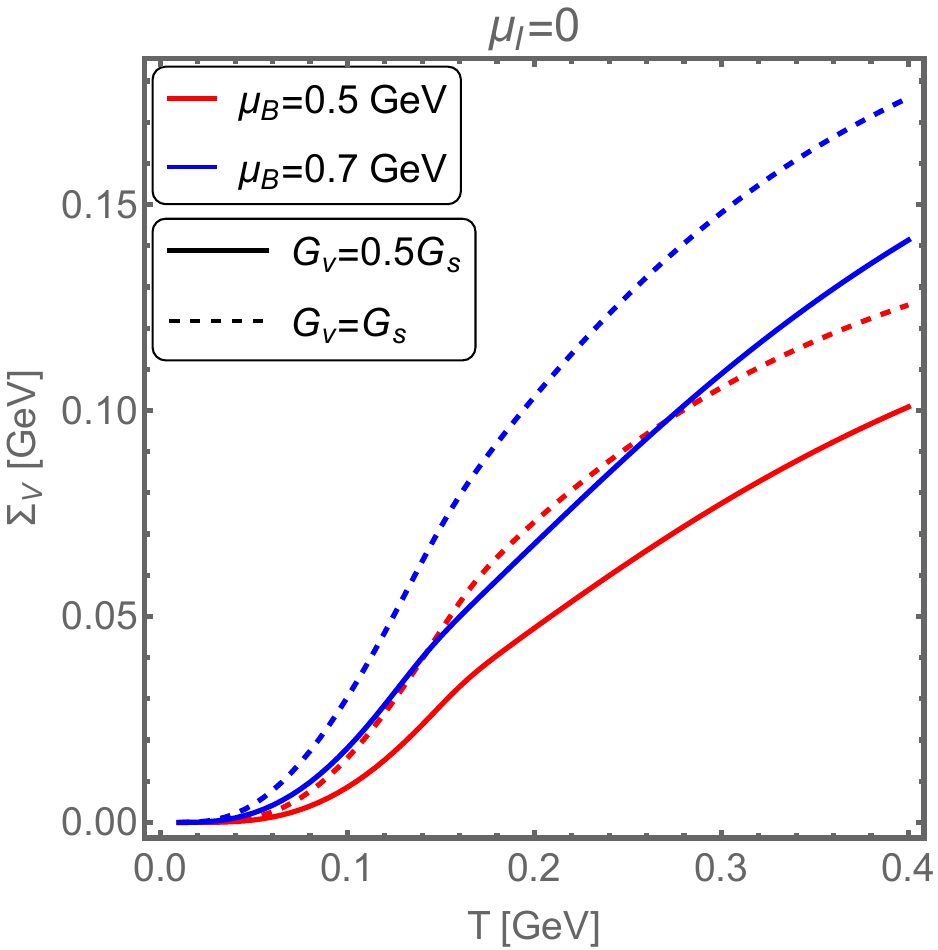}
    \hspace*{0.2cm}
    \includegraphics[scale=0.35]{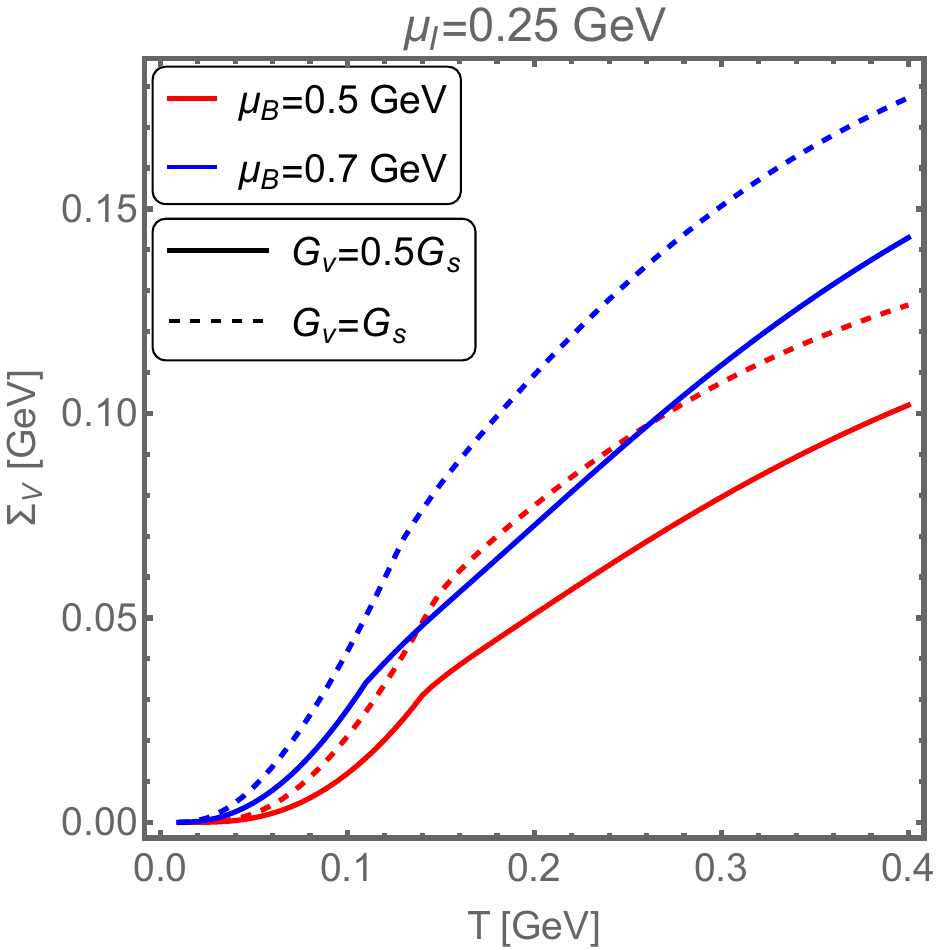}
    \hspace*{0.2cm}
    \includegraphics[scale=0.35]{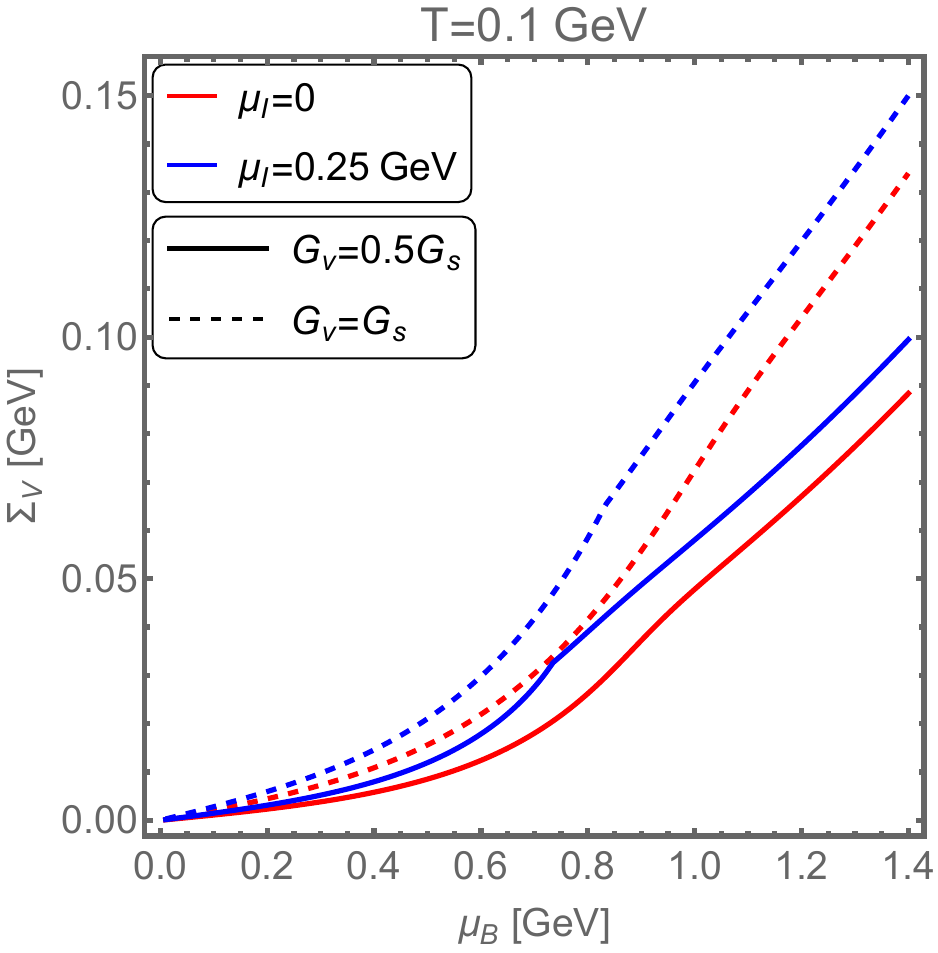}
    \caption{The vector condensate $\Sigma_V$ as a function of temperature $T$ (left and middle panels) and baryon chemical potential $\mu_B$ (right panel) within the SU(2) NJL model. Since $G_V = 0$ yields a vanishing vector condensate by definition, only $G_V = 0.5\,G_S$ and $G_V = G_S$ are considered, represented by solid and dashed lines, respectively. The left and middle panels correspond to $\mu_I = 0$ and $\mu_I = 0.25$~GeV, respectively, with red and blue curves corresponding to $\mu_B = 0.5$~GeV and $0.7$~GeV. In the right panel, red and blue curves correspond to $\mu_I = 0$ and $\mu_I = 0.25$~GeV, at fixed $T = 0.1$~GeV.}
    \label{fig:Sigma_muImuBGv}
\end{figure*}

Figure~\ref{fig:Delta_muImuBGv} shows the behavior of the pion condensate $\Delta$ as a function of $T$ and $\mu_I$ within the SU(2) NJL model. For all cases, the pion condensate decreases monotonically with increasing temperature and vanishes at a critical temperature, signaling the melting of the pion-superfluid phase. At fixed $G_V$, increasing $\mu_B$ shifts the critical temperature to lower values and reduces the temperature range where the condensate is nonzero, indicating that baryon density disfavors pion condensation. This suppression becomes more pronounced at lower values of $G_V$.  

The onset of pion condensation in the lower-right panel occurs around $\mu_I \simeq m_\pi$; at $T=0$, this happens exactly at $\mu_I = m_\pi$ due to the Silver Blaze property~\cite{Son:2000xc,Kogut:2001id,Kogut:2002zg}. For a fixed value of $\mu_B$, increasing $G_V$ shifts the onset to lower values of $\mu_I$ and generally enhances the magnitude of the condensate at fixed $\mu_I$, since the repulsive vector interaction reduces the effective baryon chemical potential and thus partially counteracts the suppression of the pion condensate by baryon density. This competing behavior, which is evidently a temperature dependent phenomenon, is most evident at $\mu_B = 0.7~\mathrm{GeV}$: at this baryon density, $\Sigma_V$ generated from a vanishing $G_V$ is insufficient to reduce the effective chemical potential $\tilde{\mu}$ enough to counteract the baryon-density-induced suppression, and the pion condensate vanishes entirely for all values of $\mu_I$ at $T = 0.1~\mathrm{GeV}$.

Hence, Figure~\ref{fig:Delta_muImuBGv} demonstrates that while an increase in baryon density tends to suppress pion condensation, the inclusion of vector interactions acts in the opposite direction. This competition nontrivially modifies both the critical temperature and the onset of pion condensation in an isospin-asymmetric matter within the SU(2) NJL framework.

Figure~\ref{fig:Sigma_muImuBGv} shows the behavior of the vector condensate $\Sigma_V$ as a function of $T$ and $\mu_B$ within the SU(2) NJL model. The vector condensate increases monotonically with temperature in the considered range, with larger values obtained for stronger vector coupling and higher baryon chemical potential‌, as evident from all three panels. Comparing the left and middle panels of Figure~\ref{fig:Sigma_muImuBGv}, one can conclude that a finite isospin chemical potential enhances the magnitude of $\Sigma_V$ at fixed $T$ and $\mu_B$, which becomes even clearer from the right panel. This indicates that isospin asymmetry indirectly strengthens the vector condensate through modifications of the quark density.

\begin{figure*}[t]
    \centering
    \includegraphics[scale=0.35]{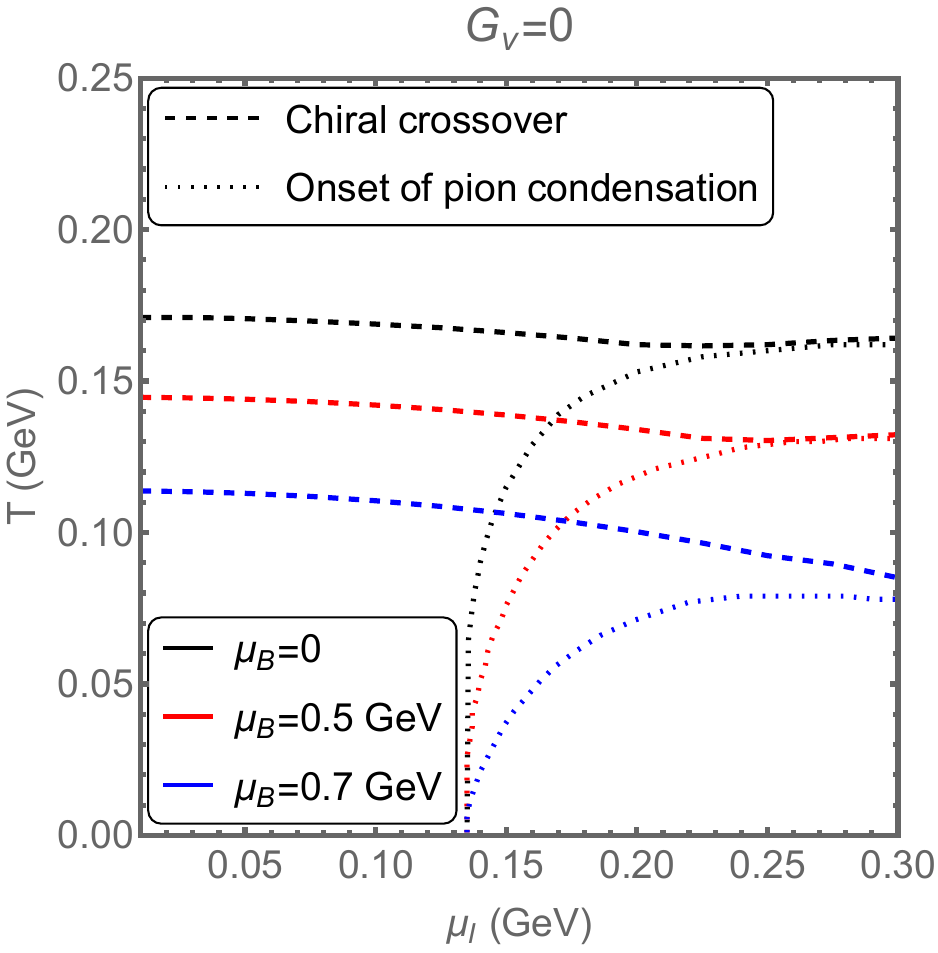} \hspace{0.2cm}\includegraphics[scale=0.35]{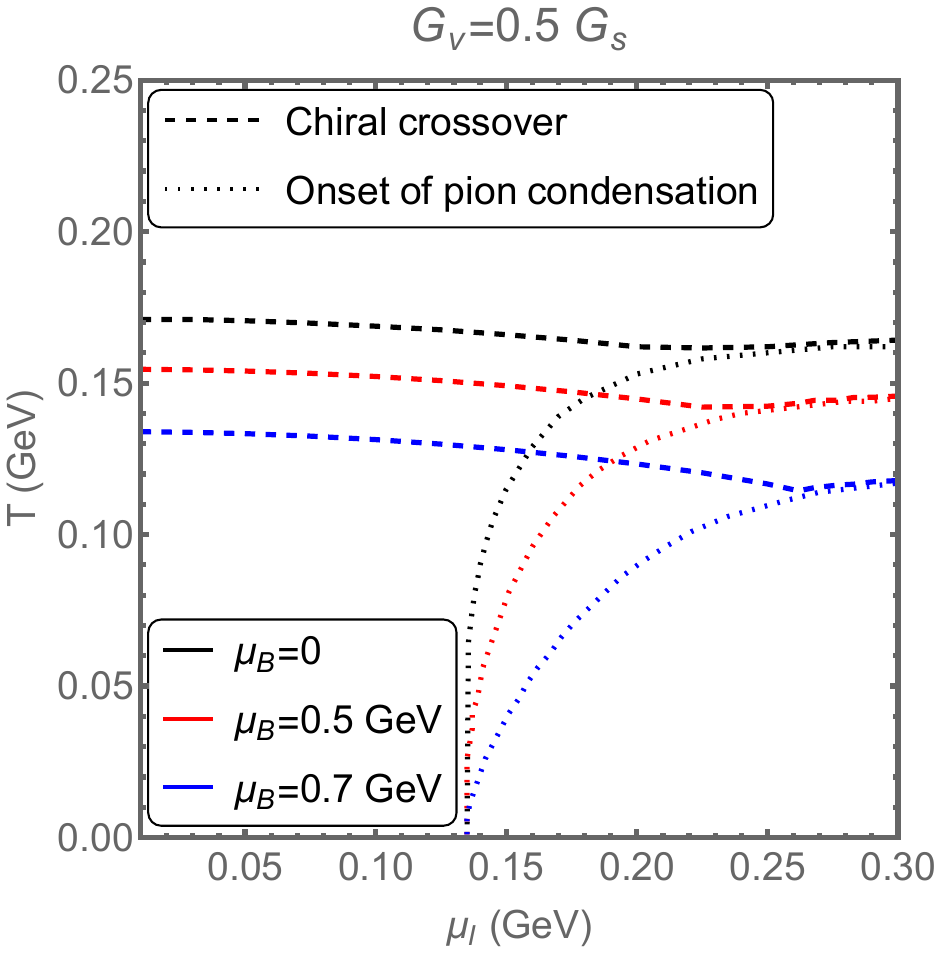} \hspace{0.2cm}
    \includegraphics[scale=0.35]{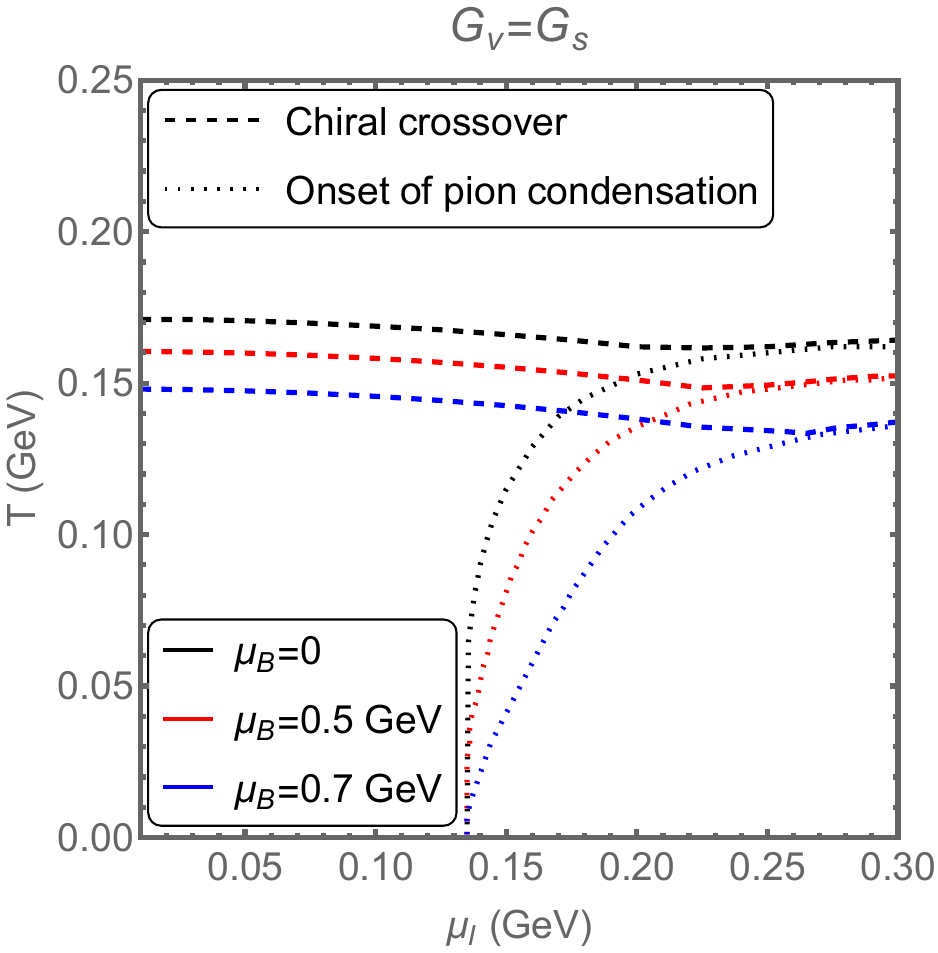} 
    \caption{The $T$--$\mu_I$ phase diagram for three values of the isoscalar-vector coupling: $G_V = 0$ (left), $G_V = 0.5\,G_S$ (middle), and $G_V = G_S$ (right). In each panel, dashed and dotted lines denote the chiral crossover boundary and the pion condensation phase boundary, respectively, for $\mu_B = 0$, $0.5$~GeV, and $0.7$~GeV, represented by black, red, and blue curves, respectively. The $\mu_B = 0$ curves (black dashed and dotted lines) are identical across all three panels.}
    \label{fig:Phase_T-muI}
\end{figure*}

\subsection{Phase diagram}

Figure~\ref{fig:Phase_T-muI} shows the $T$--$\mu_I$ phase diagram for several fixed values of $\mu_B$ and $G_V$. Both the chiral crossover and pion condensation boundaries shift to lower temperatures with increasing $\mu_B$, reflecting the suppression of both transitions by baryon density. The pion condensation boundary sets on at $\mu_I = m_\pi$ at vanishing temperatures, lies below the chiral crossover at low $\mu_I$, eventually merging with it at higher values of $\mu_I$. This carves out a distinct region of the phase diagram accessible only at sufficiently large isospin density, which we refer to as the pion condensed phase. 

\begin{figure}
    \centering
    \includegraphics[scale=0.4]{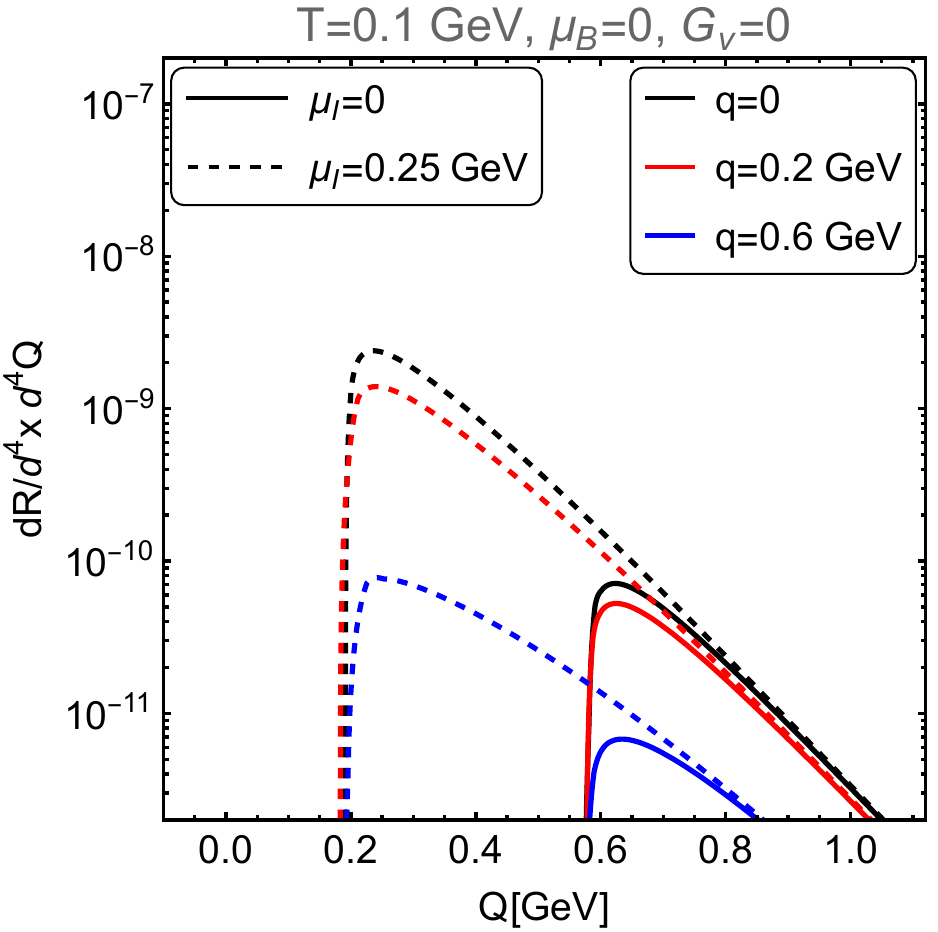}
    \caption{DPR as a function of the invariant mass $Q$ at fixed $T = 0.1$~GeV, $\mu_B = 0$, and $G_V = 0$. Black, red, and blue curves correspond to $q = 0$, $0.2$~GeV, and $0.6$~GeV, respectively, while solid and dashed curves correspond to the uncondensed ($\mu_I = 0$) and pion-condensed ($\mu_I = 0.25$~GeV) phases, respectively.}
    \label{fig:dpr_muI_wq}
\end{figure}

\begin{figure*}
    \centering
    \includegraphics[scale=0.35]{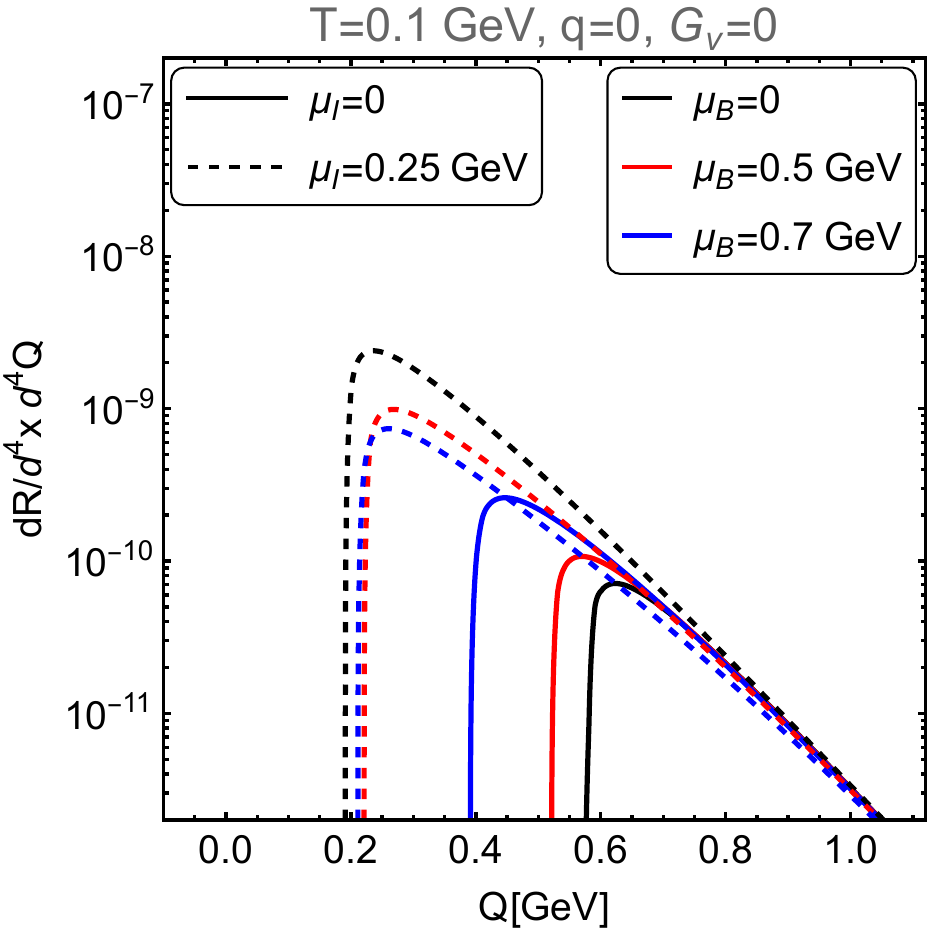}
    \hspace{0.2cm}
    \includegraphics[scale=0.35]{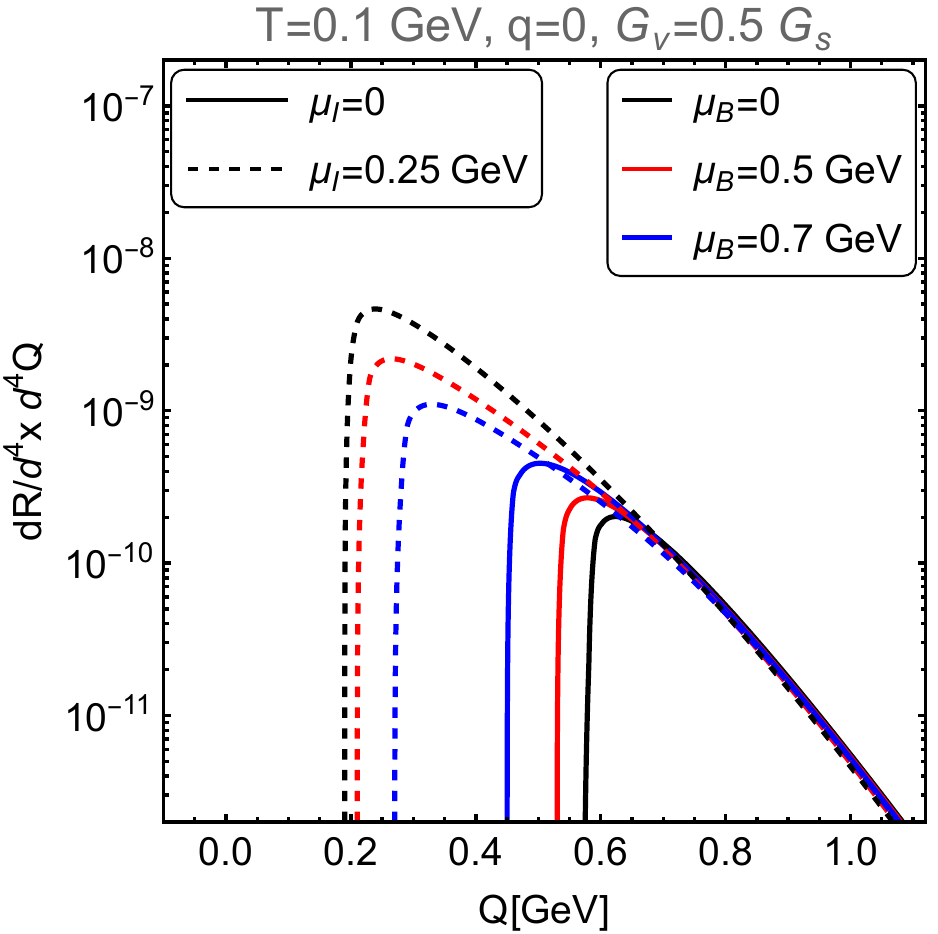}
    \hspace{0.2cm}
    \includegraphics[scale=0.35]{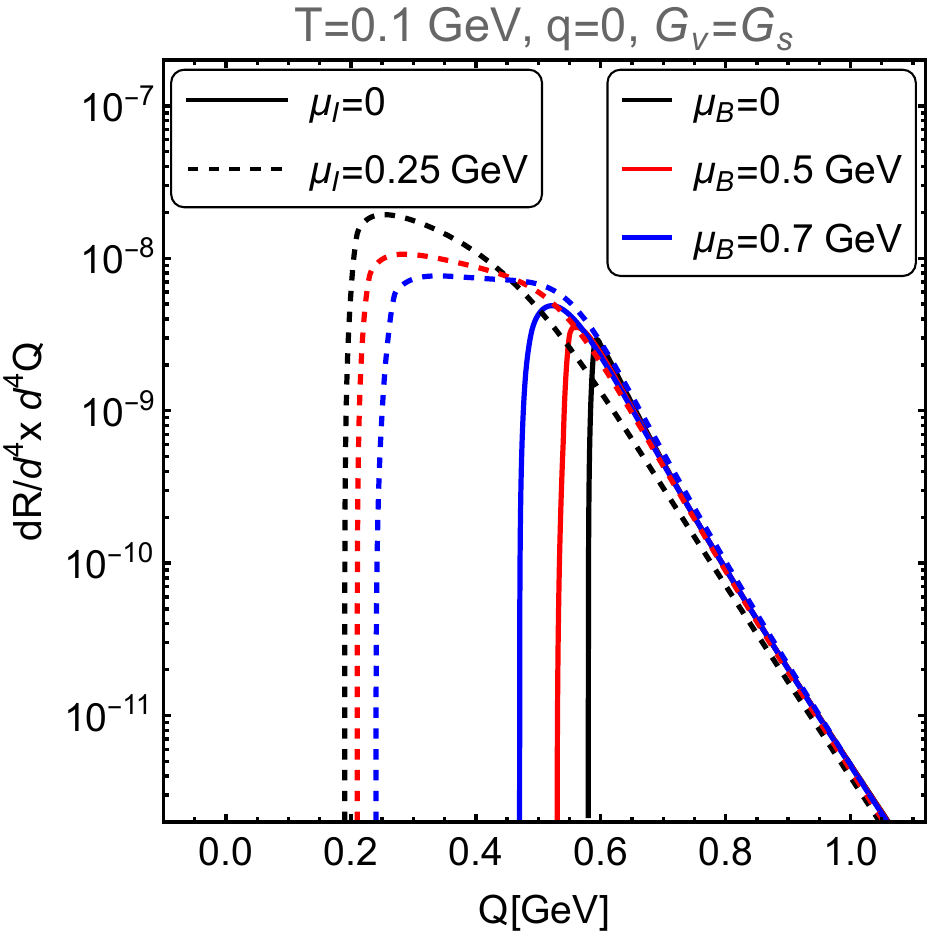}
    \caption{DPR as a function of the invariant mass $Q$ at fixed $T = 0.1$~GeV and $q = 0$, for three values of the isoscalar-vector coupling: $G_V = 0$ (left), $G_V = 0.5\,G_S$ (middle), and $G_V = G_S$ (right). In each panel, black, red, and blue curves correspond to $\mu_B = 0$, $0.5$~GeV, and $0.7$~GeV, respectively, while solid and dashed curves correspond to the uncondensed ($\mu_I = 0$) and pion-condensed ($\mu_I = 0.25$~GeV) phases, respectively.}
    \label{fig:dpr_muImuBGv1}
\end{figure*} 

At finite $G_V$, a competing effect emerges with increasing $\mu_B$: the vector mean field reduces the effective chemical potential $\tilde{\mu}$, counteracting the baryon-density-induced suppression, and shifting both the chiral crossover and pion condensation boundaries to relatively higher temperatures compared to the corresponding $G_V = 0$ curves. This is consistent with the order parameter behavior discussed in the previous subsection. The qualitative features of the $T$--$\mu_I$ phase diagram, including the onset of pion condensation at $\mu_I = m_\pi$ and the shape of the condensation boundary, are in good agreement with previous NJL model and lattice QCD studies~\cite{Lopes:2021tro,Brandt:2017oyy,Brandt:2018bwq}.

\subsection{Dilepton production rate}

\begin{figure*}
    \centering
    \includegraphics[scale=0.35]{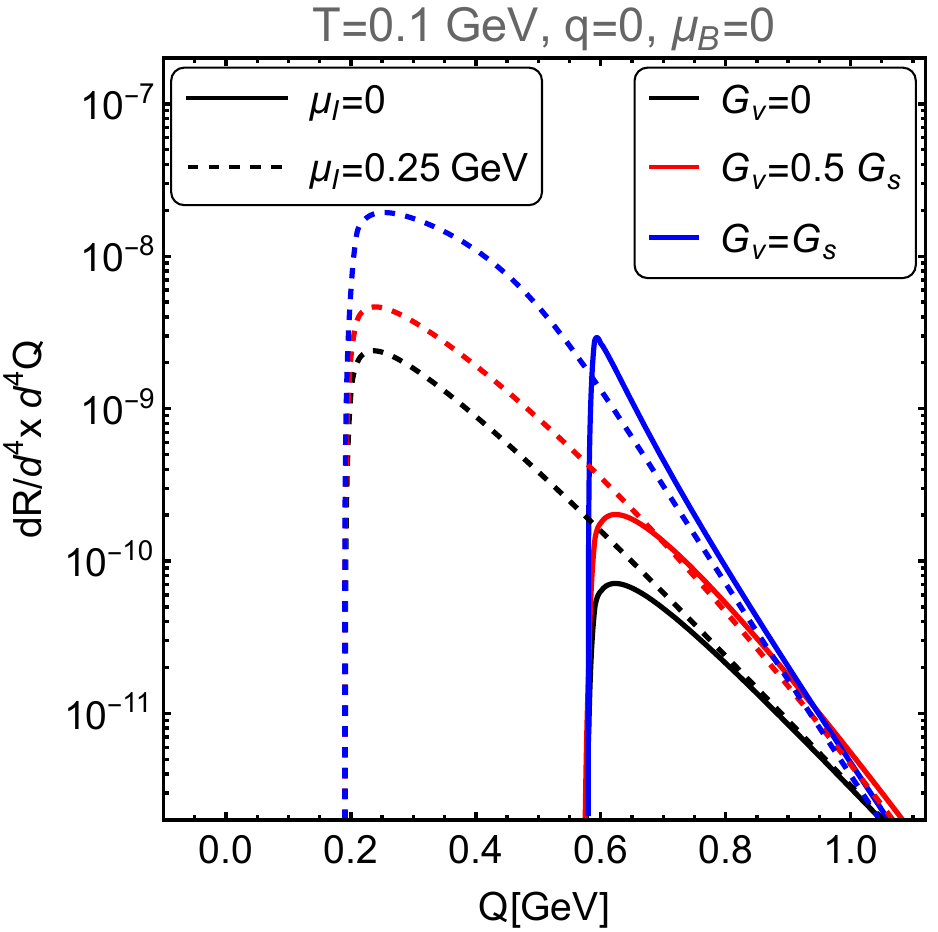}\hspace{0.2cm}
    \includegraphics[scale=0.35]{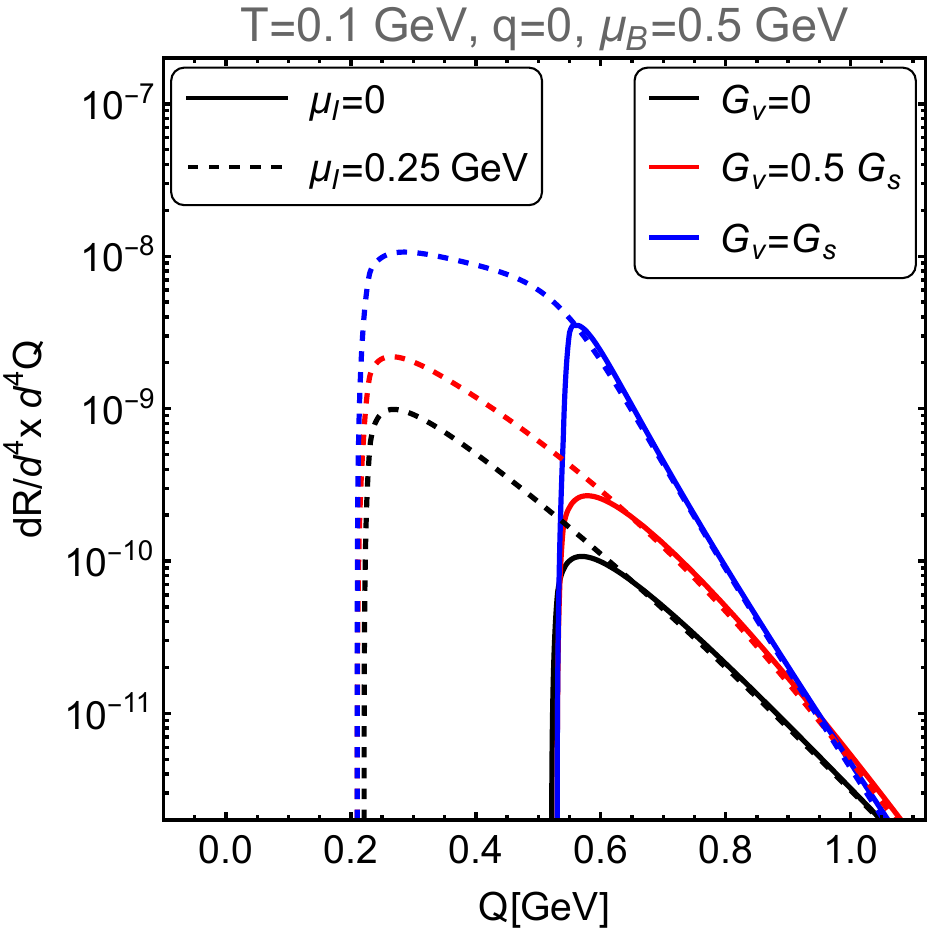}\hspace{0.2cm}
    \includegraphics[scale=0.35]{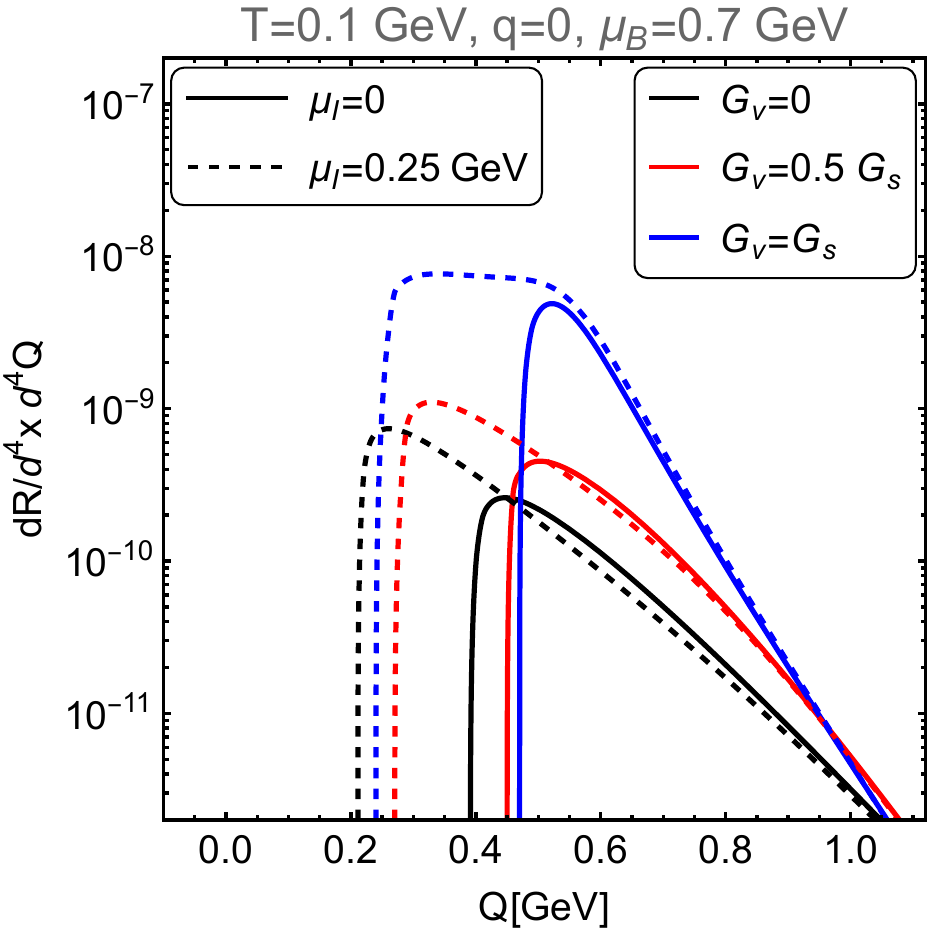}
    \caption{Complementary to Fig.~\ref{fig:dpr_muImuBGv1}, with panels organized by $\mu_B = 0$ (left), $0.5$~GeV (middle), and $0.7$~GeV (right), and curves by $G_V$, at fixed $T = 0.1$~GeV and $q = 0$. In each panel, black, red, and blue curves correspond to $G_V = 0$, $0.5\,G_S$, and $G_S$, respectively, while solid and dashed curves correspond to the uncondensed ($\mu_I = 0$) and pion-condensed ($\mu_I = 0.25$~GeV) phases, respectively.}
    \label{fig:dpr_muImuBGv2}
\end{figure*}

We begin this subsection with Figs.~\ref{fig:dpr_muI_wq},~\ref{fig:dpr_muImuBGv1}, 
and~\ref{fig:dpr_muImuBGv2}, which show the DPR as a function of the invariant mass $Q$ at fixed $T = 0.1$~GeV for different values of $\mu_I$, $\mu_B$, and $G_V$. A common feature observed in all three figures is a considerable enhancement of the DPR in the pion-condensed phase, where the onset of the dilepton rate shifts toward lower invariant masses. This enhancement primarily 
originates from the reduction of the kinematic threshold determined by the medium-dependent effective quark mass, which becomes significantly smaller in the pion-condensed phase compared to the uncondensed phase. This key mechanism underlies the sensitivity of the dilepton observable to the phase structure discussed throughout the remainder of this subsection.

In Fig.~\ref{fig:dpr_muI_wq}, the DPR is shown for vanishing baryon density and vector coupling strength, for three different values of the dilepton three-momentum $q$. Compared to the zero-momentum case, the presence of finite momentum reduces the overall magnitude of the dilepton rate. The dependence on the isospin chemical potential remains clearly visible, indicating that the medium asymmetry continues to influence the dilepton emission rate even at finite momentum.

Figure~\ref{fig:dpr_muImuBGv1} illustrates the impact of pion condensation on the DPR for different strengths of the isoscalar--vector interaction. In the absence of the isoscalar--vector interaction ($G_V = 0$; left panel), the dilepton rate displays a distinct threshold structure. The shift in the rate from the uncondensed phase ($\mu_I = 0$) to the pion-condensed phase ($\mu_I = 0.25~\mathrm{GeV}$) decreases with increasing $\mu_B$. For $\mu_B = 0.7~\mathrm{GeV}$, one observes that beyond a certain invariant mass $Q$, the contribution from the uncondensed phase surpasses that of the pion-condensed phase. This crossover occurs because the larger effective quark mass in the uncondensed phase shifts the spectral weight toward higher invariant masses, while the pion-condensed phase, with its reduced quark mass, concentrates yield predominantly in the low-$Q$ region.

When the vector interaction is introduced ($G_V = 0.5~G_S$; central panel and $G_V = G_S$; right panel), across all values of $\mu_B$ the dilepton rates receive additional enhancements in the low invariant mass region relative to the $G_V = 0$ case, and the spectral distribution is modified accordingly, as already observed in Ref.~\cite{Islam:2014sea}. At the strongest coupling considered, i.e. $G_V = G_S$ (right panel), this enhancement evolves into a pronounced plateau-like structure exclusively in the pion-condensed phase ($\mu_I = 0.25$~GeV, dashed curves), whereas the uncondensed phase ($\mu_I = 0$, solid curves) continues to exhibit the standard threshold behavior. This plateau reflects a qualitative reorganization of spectral strength at low invariant masses, driven by the interplay between the reduced effective quark mass in the pion-condensed phase and the strong resummation of the isoscalar-vector interaction through the RPA, while the uncondensed phase, retaining a larger effective quark mass, does not experience this reorganization.

\begin{figure*}[t]
    \centering
    \includegraphics[scale=0.4]{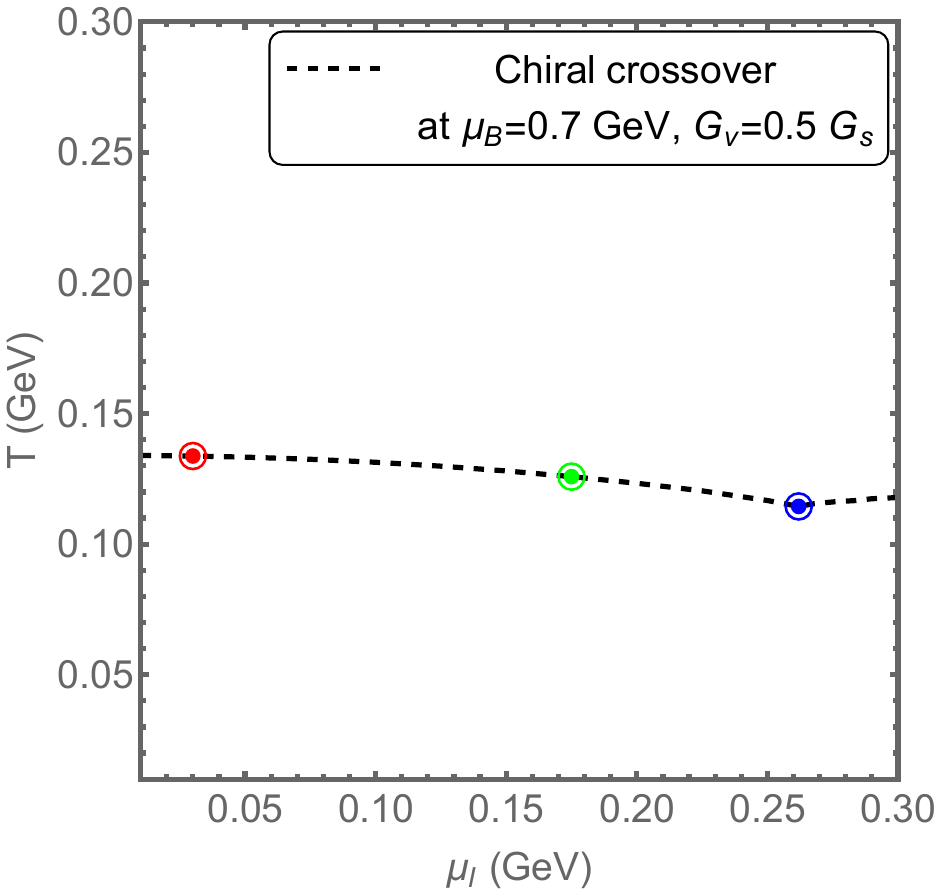}\hspace{1cm}
    \includegraphics[scale=0.4]{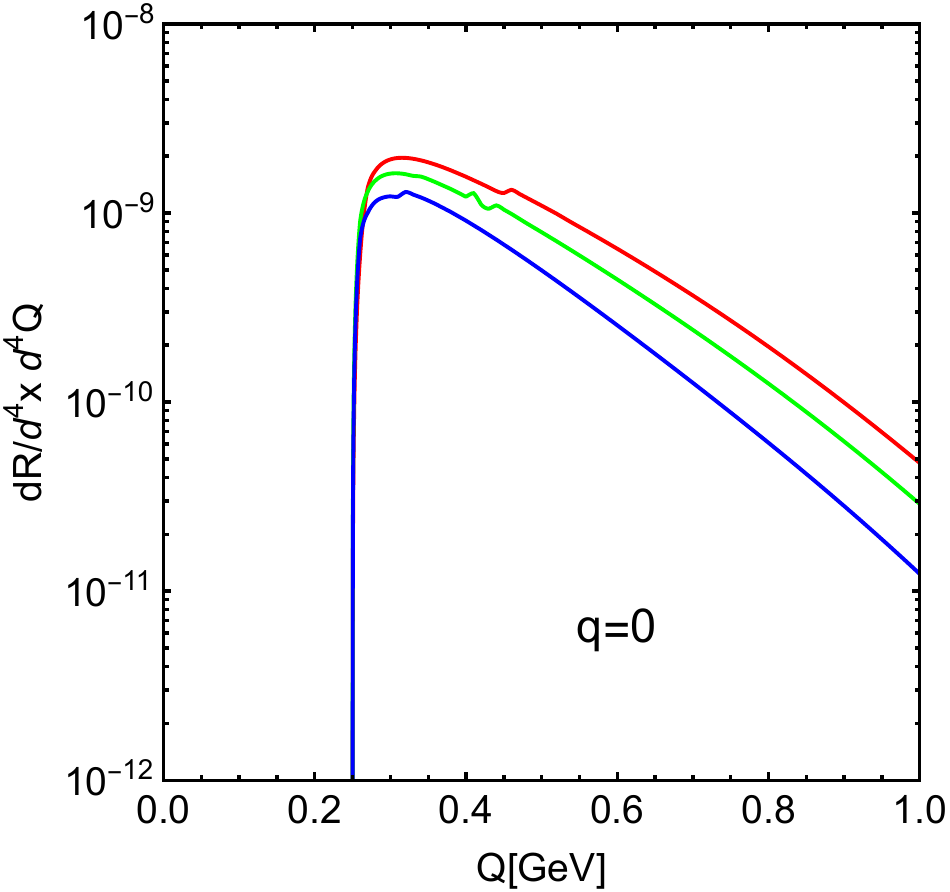}\\
    \vspace{0.5cm}
    \includegraphics[scale=0.4]{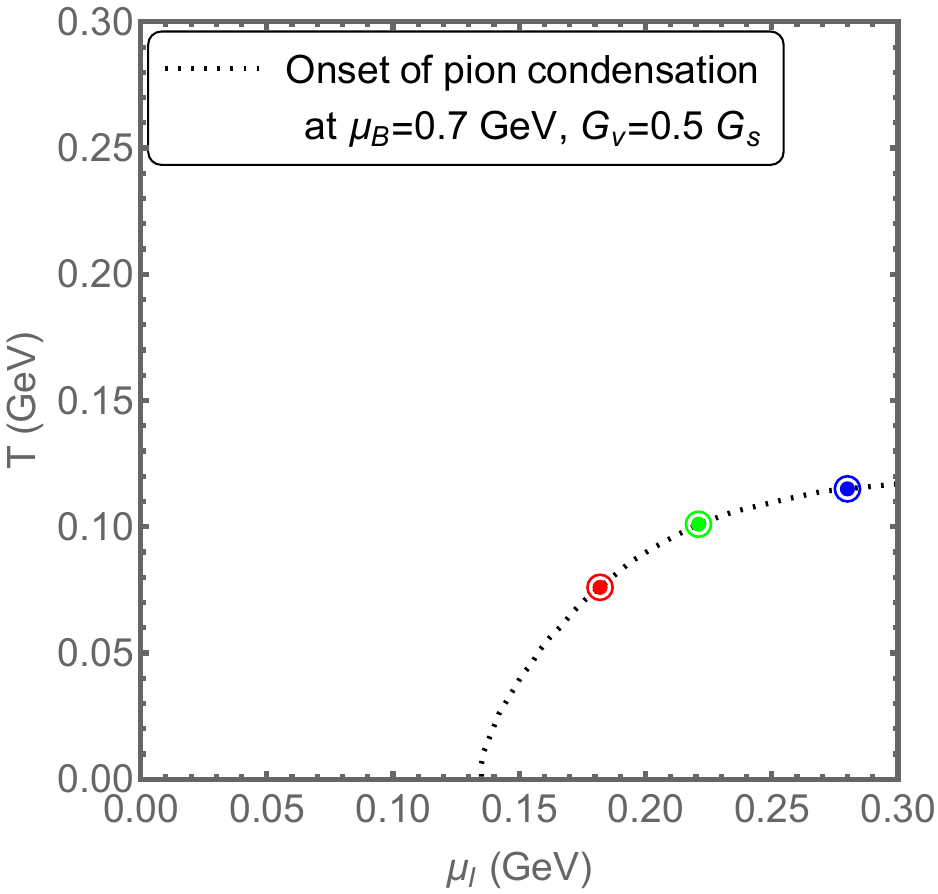}\hspace{1cm}
    \includegraphics[scale=0.4]{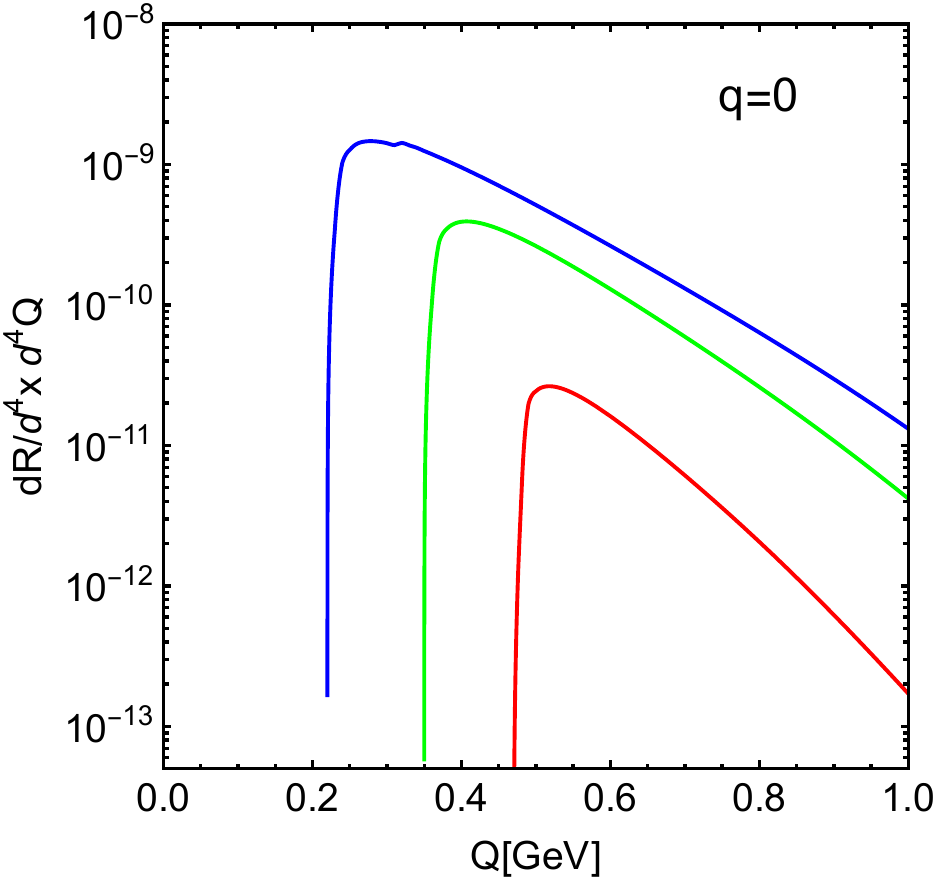}
    \caption{DPR as a function of the invariant mass $Q$ at $q = 0$, evaluated along the chiral crossover (upper panels) and pion condensation phase boundary (lower panels) of the $T$--$\mu_I$ phase diagram at fixed $\{\mu_B, G_V\} = \{0.7~\text{GeV}, 0.5\,G_S\}$. In the left panels, red, green, and blue points mark three representative points along each phase boundary, with the corresponding DPR curves in the right panels inheriting the same color coding. The coordinates along the chiral crossover boundary are $(\mu_I, T) = (0.03, 0.134)$, $(0.175, 0.126)$, $(0.262, 0.114)$~GeV, and along the pion condensation boundary are $(0.182, 0.076)$, $(0.221, 0.101)$, $(0.28, 0.115)$~GeV.}
    \label{fig:dpr_TmuI_phase_bdy}
\end{figure*}

The competing effects arising from the inclusion of $G_V$, discussed in the context of Fig.~\ref{fig:dpr_muImuBGv1}, are further reflected in Fig.~\ref{fig:dpr_muImuBGv2}. In this figure, we present the combined effects of finite isospin density and the isoscalar--vector interaction on the DPR for $\mu_B = 0$, $0.5$~GeV, and $0.7$~GeV. The $\mu_B = 0$ panel (left) confirms that a hint of flattening in the pion-condensed phase is already visible (for $G_V = G_S$) in the absence of baryon density and therefore cannot be attributed solely to baryon-density-induced modifications of the effective chemical potential. Increasing the vector coupling from $G_V = 0$ to $G_V = G_S$ leads to an enhancement of the dilepton yield for all values of the baryon chemical potential, primarily attributable to the resummed vector current correlator incorporating the additional isoscalar--vector four-fermion interactions. The progressive flattening of the dilepton rate exclusively in the pion-condensed phase across all values of $\mu_B$ is also clearly visible. This robustness establishes the plateau as a distinctive consequence of the interplay between pion condensation and the strong isoscalar-vector interaction.

Both the modification of the production thresholds and the plateau like structure in the pion condensed phase at $G_V = G_S$ become more visible at $\mu_B = 0.7$~GeV, where the effective quark chemical potential plays an important role. These results indicate that the interplay between isospin asymmetry, pion condensation, and isoscalar--vector interactions can significantly affect the DPR in a hot and dense strongly interacting matter, with the plateau structure at strong vector coupling serving as a potentially observable signature of the
pion-condensed phase.

\begin{figure*}[t]
    \centering
    \includegraphics[scale=0.4]{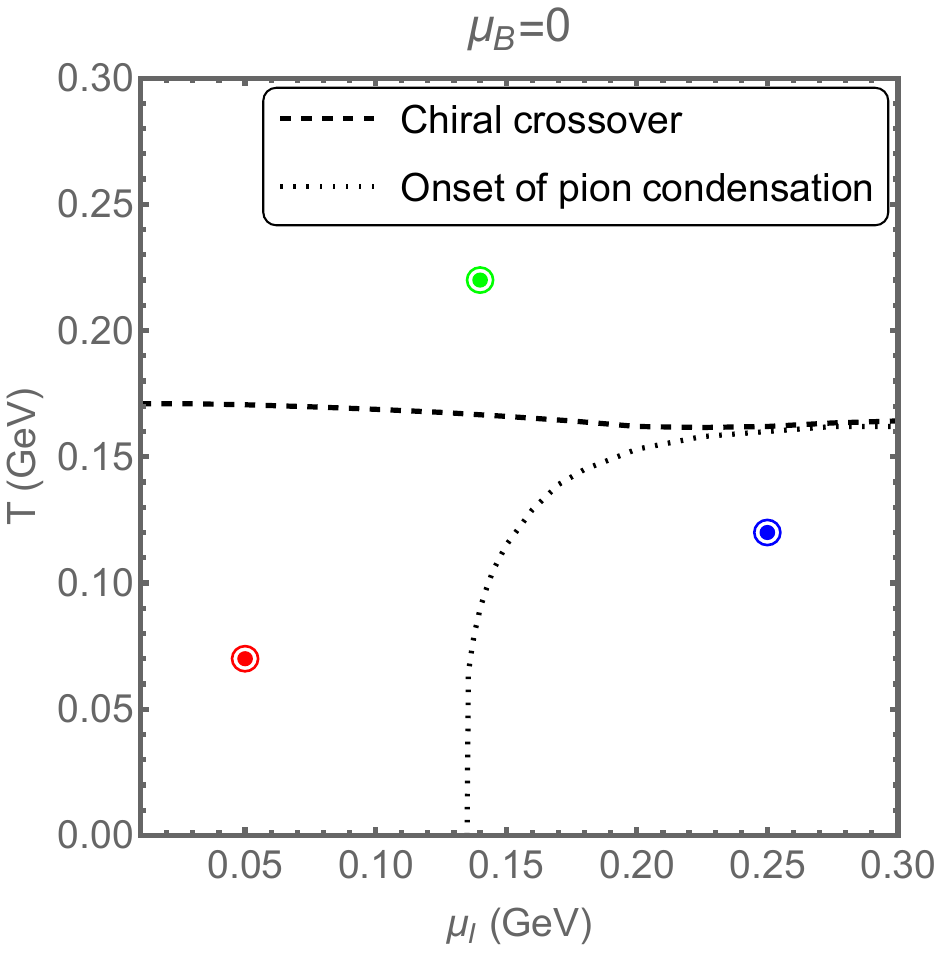}
    \includegraphics[scale=0.4]{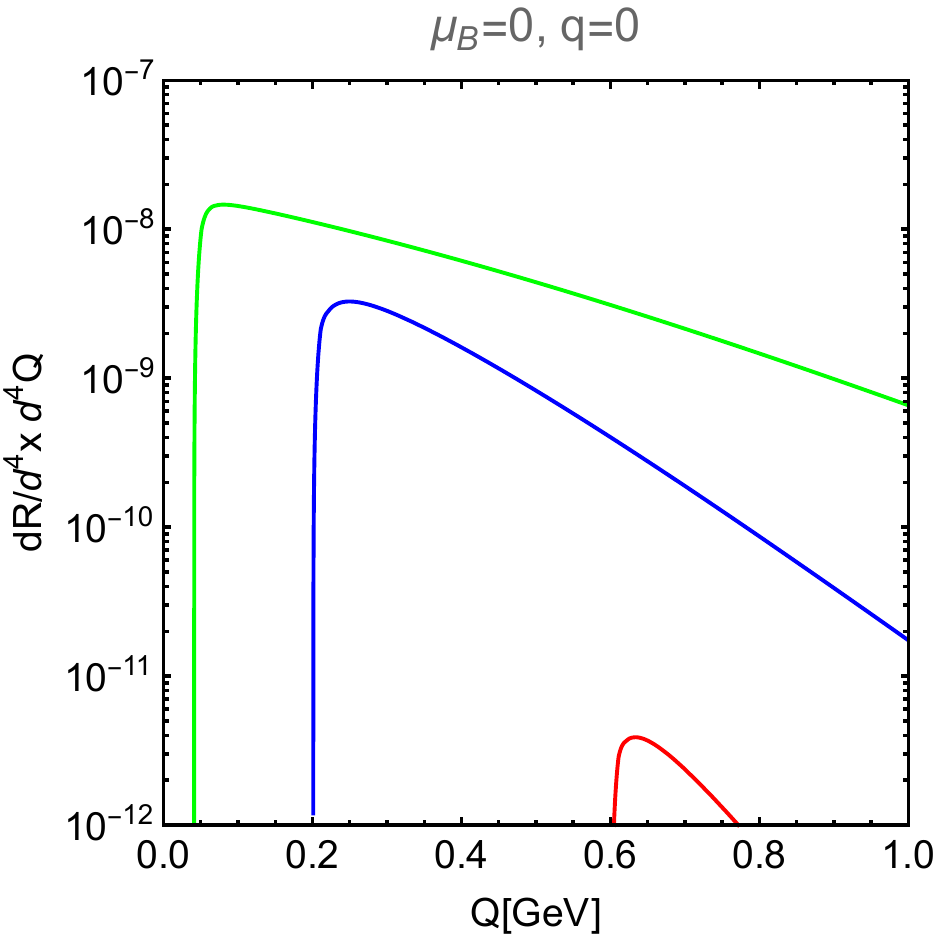}\\
    \includegraphics[scale=0.4]{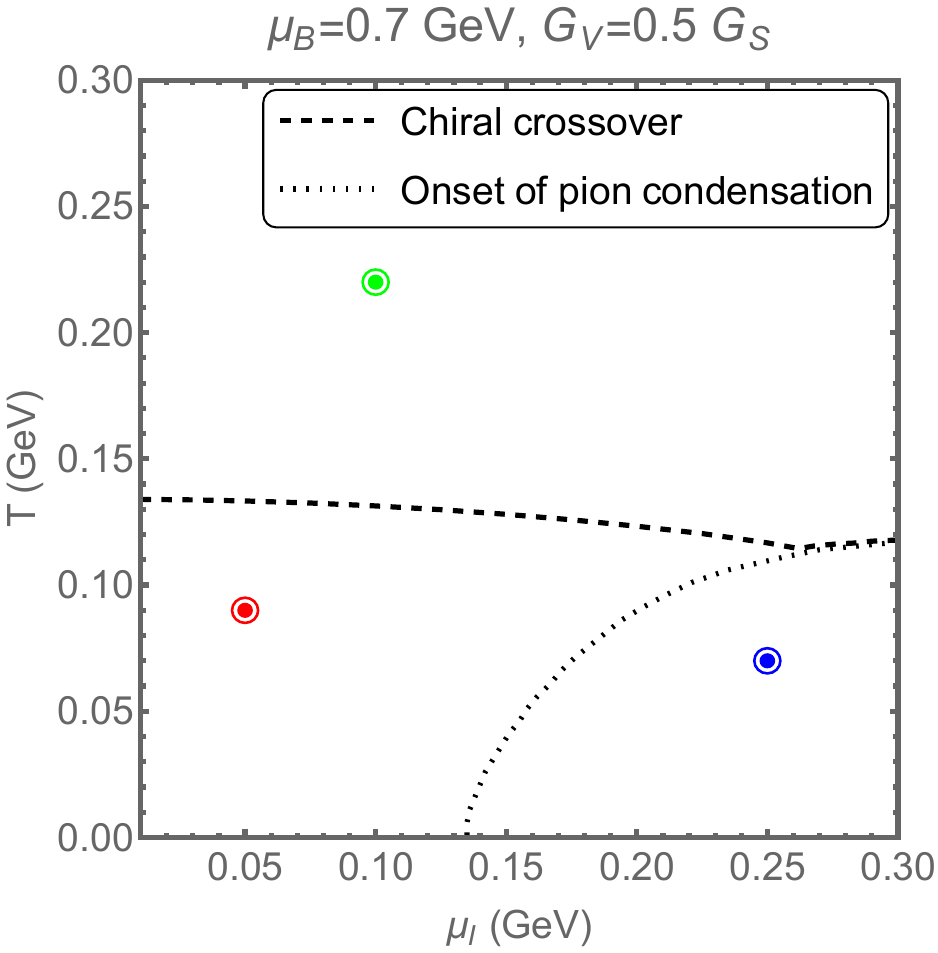}
    \includegraphics[scale=0.4]{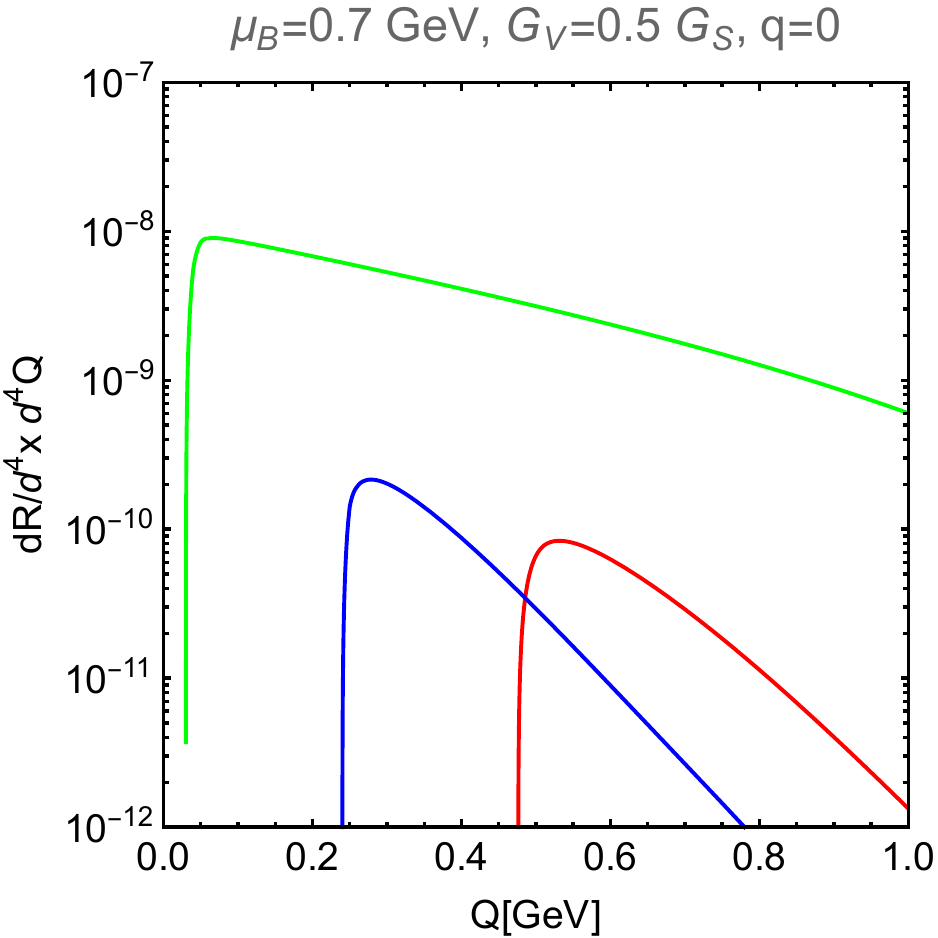}
    \caption{DPR as a function of the invariant mass $Q$ at $q = 0$, evaluated at three representative points sampling the chirally broken (red), chirally restored (green), and pion-condensed (blue) phases of the $T$--$\mu_I$ phase diagram, with the corresponding DPR curves in the right panels inheriting the same color coding. The upper (lower) panels correspond to $\mu_B = 0$ ($\mu_B = 0.7$~GeV, $G_V = 0.5\,G_S$). The chosen points are $(\mu_I, T) = (0.05, 0.07)$, $(0.14, 0.22)$, $(0.25, 0.12)$~GeV for $\mu_B = 0$, and $(0.05, 0.09)$, $(0.10, 0.22)$, $(0.25, 0.07)$~GeV for $\mu_B = 0.7$~GeV with $G_V = 0.5\,G_S$.}
    \label{fig:dpr_TmuI_diff_phases}
\end{figure*}

After discussing the general features of the DPR in the isospin asymmetric hot and dense medium within SU(2) NJL model including isoscalar-vector interactions, we illustrate how the DPR is influenced by the $T$--$\mu_I$ phase structure of the medium in Figures~\ref{fig:dpr_TmuI_phase_bdy}, and \ref{fig:dpr_TmuI_diff_phases}.

Figure~\ref{fig:dpr_TmuI_phase_bdy} tracks how the dilepton yield evolves as the system moves along the phase boundaries in the $T$--$\mu_I$ plane at fixed $\{\mu_B, G_V\} = \{0.7~\text{GeV}, 0.5\,G_S\}$. A striking contrast emerges between the two boundaries: moving along the chiral crossover produces only marginal changes in the dilepton yield, while traversing the pion-condensation boundary leads to pronounced changes in both the threshold behavior and the overall magnitude. This asymmetry establishes the dilepton rate as a considerably more sensitive probe of pion condensation than of chiral restoration.

Figure~\ref{fig:dpr_TmuI_diff_phases} provides a complementary perspective by evaluating the DPR at representative points from three distinct regions of the $T$--$\mu_I$ phase diagram. A clear hierarchy emerges in the DPR: the chirally broken phase exhibits the highest kinematic threshold, the chirally restored phase the lowest, and the pion-condensed phase an intermediate behavior, directly reflecting the ordering of the effective quark masses across the three regions. This comparison further emphasizes the sensitivity of the dilepton observable to the phase structure of strongly interacting isospin-asymmetric matter.
 
Taken together, the results of this section demonstrate that the DPR encodes a rich and distinctive dependence on the phase structure of isospin-asymmetric QCD matter. The dominant effect is the lowering of the kinematic threshold in the pion-condensed phase, which produces a robust enhancement of the low-invariant-mass dilepton yield that persists across variations in $\mu_B$, $G_V$, and the dilepton three-momentum $q$. 

The inclusion of the isoscalar-vector interaction introduces further modifications through the resummed vector correlator: while at moderate coupling ($G_V = 0.5\,G_S$) these remain quantitative in nature, at strong coupling ($G_V = G_S$) a qualitatively new feature emerges exclusively in the pion-condensed phase — a pronounced plateau-like structure at low invariant masses that is entirely absent in the uncondensed phase and persists across all values of $\mu_B$ considered, including $\mu_B = 0$. This plateau constitutes an additional discriminating signature of pion condensation beyond the threshold shift alone. Increasing baryon density tends to suppress the overall yield by reducing the effective isospin chemical potential available for condensation, but does not wash out either of these signatures. Finally, the behavior of the dilepton yield along the phase boundaries further reinforces this picture: the strong variation of the yield along the pion-condensation boundary, in stark contrast to the near-invariance along the chiral crossover, establishes dileptons as a particularly promising observable for identifying pion-condensed phases of QCD matter.

\section{Summary and Conclusions}
\label{sec:conclusion}

In this work, we have studied dilepton production from a hot and dense strongly interacting matter in the presence of finite isospin asymmetry using a two-flavor NJL model. Dileptons, as electromagnetic probes, provide direct access to the in-medium spectral properties of quarks and mesons, escaping the medium with minimal final-state interactions. Our analysis incorporates scalar, pseudoscalar, and isoscalar-vector interaction channels. The scalar and pseudoscalar interactions govern dynamical quark mass generation and pion condensation, while the isoscalar-vector interaction modifies the effective quark chemical potential and enhances the vector condensate, impacting the electromagnetic spectral function. The DPR is computed via a RPA resummation of the vector current correlator, systematically including medium-modified quark masses, effective chemical potentials, and isospin asymmetry effects.

Regarding the behavior of the order parameters, the effective quark mass $M$ exhibits a nontrivial dependence on $\mu_I$, $\mu_B$, and $G_V$, with finite isospin asymmetry leading to a significant reduction of $M$ in the vicinity of the chiral transition compared to the isospin-symmetric case. The pion condensate $\Delta$ decreases with temperature and vanishes at a critical temperature that is sensitive to the values of $\mu_B$ and $G_V$. Although increasing baryon density suppresses pion condensation by reducing the effective isospin chemical potential available for condensation, the isoscalar-vector interaction opposes this suppression by shifting the effective quark chemical potential, thus stabilizing the pion-condensed phase and shifting its onset to lower values of $\mu_I$ while enhancing the magnitude of the condensate. The vector condensate $\Sigma_V$ grows with temperature, baryon density, and vector coupling, with finite $\mu_I$ providing an additional enhancement through modifications of the quark number density.

The $T$--$\mu_I$ phase diagram reveals a clear interplay between the chiral crossover and pion condensation boundaries. Increasing $G_V$ counters the effect of finite $\mu_B$, effectively stabilizing both transitions at higher temperatures. The qualitative trends of the phase boundaries agree well with previous effective model studies~\cite{Lopes:2021tro} and lattice QCD studies~\cite{Brandt:2017oyy, Brandt:2018bwq}.

Concerning the DPR, it is significantly enhanced in the pion-condensed phase due to the reduction of the effective quark mass, which lowers the kinematic threshold for dilepton emission. Finite three-momentum reduces the overall magnitude of the DPR, but does not eliminate the sensitivity to $\mu_I$. The inclusion of the isoscalar-vector interaction enhances the DPR at low invariant masses and modifies the spectral distribution, reflecting the resummed vector correlator in the medium. 

Strikingly, at strong vector coupling $G_V = G_S$, the pion-condensed phase develops a pronounced plateau-like structure at low invariant masses, a feature absent in the uncondensed phase and persistent across all values of $\mu_B$ considered, including $\mu_B = 0$. This plateau reflects a qualitative reorganization of spectral strength driven by the interplay between the reduced effective quark mass in the pion-condensed phase and the strong resummation of the isoscalar-vector interaction through the RPA, and constitutes a potentially distinctive and observable signature of pion condensation. 

The interplay between $\mu_I$, $\mu_B$, and $G_V$ leads to distinctive features in the DPR that directly correlate with the underlying phase structure, especially along the pion-condensation boundary. In particular, the evolution of the dilepton rate along the pion-condensation phase boundary shows noticeable changes in both the threshold behavior and the overall yield, in stark contrast to the chiral crossover boundary where changes are minimal, establishing the dilepton rate as a more sensitive probe of pion condensation than of chiral restoration.

Overall, our study demonstrates that dilepton emission serves as a sensitive probe of isospin asymmetry and pion condensation in strongly interacting matter. Modifications in the effective quark mass, vector condensate, and vector current correlator leave observable imprints on the dilepton yield, particularly in the low- to intermediate-invariant mass region, with possible implications for ongoing and future experimental programs at FAIR~\cite{CBM:2016kpk}, J-PARC~\cite{Sako:2014fha}, and NICA~\cite{Kekelidze:2017tgp}, where conditions of strong isospin imbalance and moderate temperatures may be realized.

The present analysis is carried out within the mean-field (Hartree) approximation of the NJL model, treating the quark propagator in the standard NJL basis. Several extensions of this work would be of considerable interest. 

First, a more complete treatment of the DPR in the pion-condensed phase would require the use of the full Nambu--Gorkov propagator, which explicitly accounts for the off-diagonal quark-antiquark pairing structure induced by the pion condensate. Such a calculation would go beyond the present framework and is expected to introduce additional contributions to the vector current correlator arising from the anomalous propagator components, potentially modifying the dilepton yield in a nontrivial manner in the deeply condensed phase. 

Second, a natural and phenomenologically important extension would be the merging of the present quark-level dilepton rate with the hadronic dilepton rate in the low invariant mass region. Since the NJL model provides a consistent description of both the quark and mesonic degrees of freedom, a unified treatment incorporating both the quark-antiquark annihilation contribution computed here and the hadronic contributions from $\pi-\pi$ annihilation and in-medium $\rho$ meson spectral functions would provide a more complete picture of dilepton emission across the full invariant mass spectrum. Furthermore, extensions to include the effects of a finite strangeness chemical potential, or beyond mean-field fluctuations would be interesting directions for future investigation.

\section{Acknowledgements}

A.~B. acknowledges support from the ULAM fellowship program of the Polish National Agency for Academic Exchange (NAWA), No.~BNI/ULM/2024/1/00193 and EU’s NextGenerationEU instrument through the National Recovery and Resilience Plan of Romania - Pillar III-C9-I8, managed by the Ministry of Research, Innovation and Digitization, within the project entitled ``Facets of Rotating Quark-Gluon Plasma'' (FORQ), contract no.~760079/23.05.2023 code CF 103/15.11.2022. The work of K.~R. and C.~S. is supported by the National Science Centre (NCN), Poland, under OPUS Grant No.~2022/45/B/ST2/01527. K.~R. acknowledges the support of the Polish Ministry of Science and Higher Education. C.~S. acknowledges the support of the World Premier International Research Center Initiative (WPI) under MEXT, Japan.

\appendix

\section{Real and Imaginary parts of the vector correlator}
\label{appA}

The imaginary part of the resummed vector current correlator for a single 
quark flavor $f$ can be expressed as :
\begin{align}
V^{\rm I}_{00;f} &=  \frac{C^{\rm I}_{00;f}}
 {\Big[ 1- 2G_V \Big(1-\frac{\omega^2}{{q}^2}\Big){C^{\rm R}_{00;f}}\Big]^2+
 \Big[2G_V (1-\frac{\omega^2}{{q}^2})
 C^{\rm I}_{00;f}\Big ]^2},\label{eq:V00} \\
V^{\rm I}_{ii;f} &= \frac{\omega^2}{{q}^2}V^{\rm I}_{00;f} \quad + \nn\\
 &\frac{C^{\rm I}_{ii;f}-\frac{\omega^2}
 {{q}^2}C^{\rm I}_{00;f}}
 {\left[1+G_V C^{\rm R}_{ii;f}-G_V
 \frac{\omega^2}
 {{q}^2}C^{\rm R}_{00;f}\right]^2
 +G_V^2\Big[C^{\rm I}_{ii;f}-\frac{\omega^2}
 {{q}^2}C^{\rm I}_{00;f}\Big]^2}, \label{eq:Vii}
\end{align}
where the superscripts ${\rm I}$ and ${\rm R}$ represent the 
imaginary and real parts of the corresponding correlators, respectively.

Real parts of the temporal and spatial components of the single-flavor 
one-loop vector current-current correlator~\cite{Islam:2014sea} can be found from 
Eq.~\eqref{eq:oneloop_Cmunu} by performing the Matsubara frequency 
sum and can be expressed in terms of the spatial 
momentum integral as :
\begin{align}
    C_{00;f}^{\rm R}(\omega,\vec q) &= C_{00}^{\rm R;v}(\omega,\vec q) + C_{00;f}^{\rm R;m}(\omega,\vec q), \\
    C_{ii;f}^{\rm R}(\omega,\vec q) &= C_{ii}^{\rm R;v}(\omega,\vec q) + C_{ii;f}^{\rm R;m}(\omega,\vec q),
\end{align}
with
\begin{multline}
C_{00}^{\rm R;v} = \frac{N_c}{4\pi^2} \int_0^\Lambda dp 
\frac{p}{2 E_p q} \Bigg[ 4 pq + 6 E_p X_-- 6 E_p X_+  \\
- Y_- \ln \left| \frac{E_p+X_- - \omega}{E_p+X_+ - \omega} \right|
+ Y_+ \ln \left| \frac{E_p+X_+ + \omega}{E_p+X_- + \omega} \right| \Bigg],
\label{eq:C00_Rv}
\end{multline}
\begin{multline}
C_{00;f}^{\rm R;m}= \frac{N_c}{2\pi^2} \int_0^\infty dp \, p \, 
\aleph_f \Bigg[ \frac{\omega}{q} \ln \left| \frac{Q^2 - 4(pq + \omega E_p)^2}{Q^2 - 4(pq - \omega E_p)^2} \right|\\
- \frac{4 E_p^2 + Q^2}{4 q E_p} \ln \left| \frac{(Q^2 - 2pq)^2 - 4 \omega^2 E_p^2}{(Q^2 + 2pq)^2 - 4 \omega^2 E_p^2} \right|- \frac{2p}{E_p} \Bigg],
\label{eq:C00_Rm}
\end{multline}
\begin{multline}
C_{ii}^{\rm R;v} = \frac{N_c}{4\pi^2} \int_0^\Lambda dp \, p \, \frac{1}{2 E_p q} \Big[10 E_p (X_- - X_+) -4 pq \\
- Z_- \ln \left| \frac{E_p+X_- - \omega}{E_p+X_+ - \omega} \right|
+ Z_+ \ln \left| \frac{E_p+X_+ + \omega}{E_p+X_- + \omega} \right| \Big],
\label{eq:Cii_Rv}
\end{multline}
\begin{multline}
C_{ii;f}^{\rm R;m}= \frac{N_c}{4\pi^2} \int_0^\infty dp \, p \,
\aleph_f \Bigg[ 2\frac{\omega}{q} \ln \left| \frac{Q^2 - 4(pq + \omega E_p)^2}{Q^2 - 4(pq - \omega E_p)^2} \right|+ \\
\frac{Q^2 - 4p^2}{2 q E_p} \ln \left| \frac{(Q^2 - 2pq)^2 - 4 \omega^2 E_p^2}{(Q^2 + 2pq)^2 - 4 \omega^2 E_p^2} \right|+ 4 \frac{p}{E_p} \Bigg].
\label{eq:Cii_Rm}
\end{multline}
Here, we have separated the vacuum and medium parts by superscripts ${\rm v}$ 
and ${\rm m}$ with an ultraviolet cutoff $\Lambda$ appearing in the momentum 
integration for the vacuum terms; $\aleph_f = f(E_p - \tilde{\mu}_f) + 
f(E_p + \tilde{\mu}_f)$ is the single-flavor thermal distribution with $f(E_p \pm \tilde{\mu}_f)=1/(1+e^{\beta(E_p\pm\tilde\mu_f)})$ and
$\tilde{\mu}_f = \mu_B/3 - \Sigma_V + I_f\,\mu_I/2$ ($I_f = +1$ for $u$, 
$I_f = -1$ for $d$). The other variables defined for brevity are :
\begin{multline}
X_\pm = \sqrt{E_p^2 \pm 2 pq + q^2}, \quad
Y_\pm = 4 E_p^2 \pm 4 E_p \omega + Q^2, \quad \\
Z_\pm = 4 p^2 \pm 4 E_p \omega - Q^2, \quad
E_p = \sqrt{p^2 + M^2}.
\end{multline}

For the evaluation of the imaginary parts of $C_{00}$ and $C_{ii}$, one can further restrict the kinematical domains of the contributing processes, which arises from the kinematically allowed delta functions and, in turn, provides finite limits to the spatial momentum integration. The corresponding combined (vacuum and medium) final expressions for a single quark flavor 
$f$ are~\cite{Islam:2014sea}
\begin{multline}
C^{\rm I}_{00;f}(\omega,\vec q) = \frac{N_c}{4\pi} \int_{p_-}^{p_+} dp \, p \, 
\frac{4 \omega E_p - 4 E_p^2 - Q^2}{2 E_p q} \, 
\big[\aleph_f - 1\big],
\label{eq:C00_I}
\end{multline}
\begin{multline}
 C^{\rm I}_{ii;f}(\omega,\vec q) = \frac{N_c}{4\pi} \int_{p_-}^{p_+} dp \, p \,
\frac{4 \omega E_p - 4 p^2 + Q^2}{2 E_p q} \,
\big[\aleph_f - 1\big],
\label{eq:Cii_I}
\end{multline}
with
\begin{equation}
p_\pm = \frac{\omega}{2} \sqrt{1 - \frac{4 M^2}{Q^2}} \pm \frac{q}{2}.
\end{equation}
The full DPR is then obtained by weighting the single-flavor 
spectral function $\rho_V^f$ with the corresponding electric charge squared 
and summing over flavors as in Eq.~\eqref{eq:dr_expr_nonzeromuI}. In the 
isospin symmetric limit $\mu_I \to 0$, one recovers $\tilde{\mu}_u = 
\tilde{\mu}_d \equiv \tilde{\mu}$, and 
$\sum_{f}q_f^2\,\rho_V^f = \frac{5}{9}\,e^2\,\rho_V$, consistently reducing to 
Eq.~\eqref{eq:dr_expr}.


\begin{thebibliography}{63}%
\makeatletter
\providecommand \@ifxundefined [1]{%
 \@ifx{#1\undefined}
}%
\providecommand \@ifnum [1]{%
 \ifnum #1\expandafter \@firstoftwo
 \else \expandafter \@secondoftwo
 \fi
}%
\providecommand \@ifx [1]{%
 \ifx #1\expandafter \@firstoftwo
 \else \expandafter \@secondoftwo
 \fi
}%
\providecommand \natexlab [1]{#1}%
\providecommand \enquote  [1]{``#1''}%
\providecommand \bibnamefont  [1]{#1}%
\providecommand \bibfnamefont [1]{#1}%
\providecommand \citenamefont [1]{#1}%
\providecommand \href@noop [0]{\@secondoftwo}%
\providecommand \href [0]{\begingroup \@sanitize@url \@href}%
\providecommand \@href[1]{\@@startlink{#1}\@@href}%
\providecommand \@@href[1]{\endgroup#1\@@endlink}%
\providecommand \@sanitize@url [0]{\catcode `\\12\catcode `\$12\catcode
  `\&12\catcode `\#12\catcode `\^12\catcode `\_12\catcode `\%12\relax}%
\providecommand \@@startlink[1]{}%
\providecommand \@@endlink[0]{}%
\providecommand \url  [0]{\begingroup\@sanitize@url \@url }%
\providecommand \@url [1]{\endgroup\@href {#1}{\urlprefix }}%
\providecommand \urlprefix  [0]{URL }%
\providecommand \Eprint [0]{\href }%
\providecommand \doibase [0]{http://dx.doi.org/}%
\providecommand \selectlanguage [0]{\@gobble}%
\providecommand \bibinfo  [0]{\@secondoftwo}%
\providecommand \bibfield  [0]{\@secondoftwo}%
\providecommand \translation [1]{[#1]}%
\providecommand \BibitemOpen [0]{}%
\providecommand \bibitemStop [0]{}%
\providecommand \bibitemNoStop [0]{.\EOS\space}%
\providecommand \EOS [0]{\spacefactor3000\relax}%
\providecommand \BibitemShut  [1]{\csname bibitem#1\endcsname}%
\let\auto@bib@innerbib\@empty
\bibitem [{\citenamefont {Shuryak}(1978)}]{Shuryak:1978ij}%
  \BibitemOpen
  \bibfield  {author} {\bibinfo {author} {\bibfnamefont {E.~V.}\ \bibnamefont
  {Shuryak}},\ }\href {\doibase 10.1016/0370-2693(78)90370-2} {\bibfield
  {journal} {\bibinfo  {journal} {Phys. Lett. B}\ }\textbf {\bibinfo {volume}
  {78}},\ \bibinfo {pages} {150} (\bibinfo {year} {1978})}\BibitemShut
  {NoStop}%
\bibitem [{\citenamefont {Adamczyk}\ \emph {et~al.}(2014)\citenamefont
  {Adamczyk} \emph {et~al.}}]{STAR:2013pwb}%
  \BibitemOpen
  \bibfield  {author} {\bibinfo {author} {\bibfnamefont {L.}~\bibnamefont
  {Adamczyk}} \emph {et~al.} (\bibinfo {collaboration} {STAR}),\ }\href
  {\doibase 10.1103/PhysRevLett.113.022301} {\bibfield  {journal} {\bibinfo
  {journal} {Phys. Rev. Lett.}\ }\textbf {\bibinfo {volume} {113}},\ \bibinfo
  {pages} {022301} (\bibinfo {year} {2014})},\ \bibinfo {note} {[Addendum:
  Phys.Rev.Lett. 113, 049903 (2014)]},\ \Eprint
  {http://arxiv.org/abs/1312.7397} {arXiv:1312.7397 [hep-ex]} \BibitemShut
  {NoStop}%
\bibitem [{\citenamefont {Adare}\ \emph {et~al.}(2016)\citenamefont {Adare}
  \emph {et~al.}}]{PHENIX:2015vek}%
  \BibitemOpen
  \bibfield  {author} {\bibinfo {author} {\bibfnamefont {A.}~\bibnamefont
  {Adare}} \emph {et~al.} (\bibinfo {collaboration} {PHENIX}),\ }\href
  {\doibase 10.1103/PhysRevC.93.014904} {\bibfield  {journal} {\bibinfo
  {journal} {Phys. Rev. C}\ }\textbf {\bibinfo {volume} {93}},\ \bibinfo
  {pages} {014904} (\bibinfo {year} {2016})},\ \Eprint
  {http://arxiv.org/abs/1509.04667} {arXiv:1509.04667 [nucl-ex]} \BibitemShut
  {NoStop}%
\bibitem [{\citenamefont {McLerran}\ and\ \citenamefont
  {Toimela}(1985)}]{McLerran:1984ay}%
  \BibitemOpen
  \bibfield  {author} {\bibinfo {author} {\bibfnamefont {L.~D.}\ \bibnamefont
  {McLerran}}\ and\ \bibinfo {author} {\bibfnamefont {T.}~\bibnamefont
  {Toimela}},\ }\href {\doibase 10.1103/PhysRevD.31.545} {\bibfield  {journal}
  {\bibinfo  {journal} {Phys. Rev. D}\ }\textbf {\bibinfo {volume} {31}},\
  \bibinfo {pages} {545} (\bibinfo {year} {1985})}\BibitemShut {NoStop}%
\bibitem [{\citenamefont {Kajantie}\ \emph {et~al.}(1986)\citenamefont
  {Kajantie}, \citenamefont {Kapusta}, \citenamefont {McLerran},\ and\
  \citenamefont {Mekjian}}]{Kajantie:1986dh}%
  \BibitemOpen
  \bibfield  {author} {\bibinfo {author} {\bibfnamefont {K.}~\bibnamefont
  {Kajantie}}, \bibinfo {author} {\bibfnamefont {J.~I.}\ \bibnamefont
  {Kapusta}}, \bibinfo {author} {\bibfnamefont {L.~D.}\ \bibnamefont
  {McLerran}}, \ and\ \bibinfo {author} {\bibfnamefont {A.}~\bibnamefont
  {Mekjian}},\ }\href {\doibase 10.1103/PhysRevD.34.2746} {\bibfield  {journal}
  {\bibinfo  {journal} {Phys. Rev. D}\ }\textbf {\bibinfo {volume} {34}},\
  \bibinfo {pages} {2746} (\bibinfo {year} {1986})}\BibitemShut {NoStop}%
\bibitem [{\citenamefont {Weldon}(1990)}]{Weldon:1990iw}%
  \BibitemOpen
  \bibfield  {author} {\bibinfo {author} {\bibfnamefont {H.~A.}\ \bibnamefont
  {Weldon}},\ }\href {\doibase 10.1103/PhysRevD.42.2384} {\bibfield  {journal}
  {\bibinfo  {journal} {Phys. Rev. D}\ }\textbf {\bibinfo {volume} {42}},\
  \bibinfo {pages} {2384} (\bibinfo {year} {1990})}\BibitemShut {NoStop}%
\bibitem [{\citenamefont {Rapp}\ and\ \citenamefont
  {Wambach}(2000)}]{Rapp:1999ej}%
  \BibitemOpen
  \bibfield  {author} {\bibinfo {author} {\bibfnamefont {R.}~\bibnamefont
  {Rapp}}\ and\ \bibinfo {author} {\bibfnamefont {J.}~\bibnamefont {Wambach}},\
  }\href {\doibase 10.1007/0-306-47101-9_1} {\bibfield  {journal} {\bibinfo
  {journal} {Adv. Nucl. Phys.}\ }\textbf {\bibinfo {volume} {25}},\ \bibinfo
  {pages} {1} (\bibinfo {year} {2000})},\ \Eprint
  {http://arxiv.org/abs/hep-ph/9909229} {arXiv:hep-ph/9909229} \BibitemShut
  {NoStop}%
\bibitem [{\citenamefont {Rapp}\ and\ \citenamefont {van
  Hees}(2016)}]{Rapp:2014hha}%
  \BibitemOpen
  \bibfield  {author} {\bibinfo {author} {\bibfnamefont {R.}~\bibnamefont
  {Rapp}}\ and\ \bibinfo {author} {\bibfnamefont {H.}~\bibnamefont {van
  Hees}},\ }\href {\doibase 10.1016/j.physletb.2015.12.065} {\bibfield
  {journal} {\bibinfo  {journal} {Phys. Lett. B}\ }\textbf {\bibinfo {volume}
  {753}},\ \bibinfo {pages} {586} (\bibinfo {year} {2016})},\ \Eprint
  {http://arxiv.org/abs/1411.4612} {arXiv:1411.4612 [hep-ph]} \BibitemShut
  {NoStop}%
\bibitem [{\citenamefont {Arnaldi}\ \emph {et~al.}(2006)\citenamefont {Arnaldi}
  \emph {et~al.}}]{NA60:2006ymb}%
  \BibitemOpen
  \bibfield  {author} {\bibinfo {author} {\bibfnamefont {R.}~\bibnamefont
  {Arnaldi}} \emph {et~al.} (\bibinfo {collaboration} {NA60}),\ }\href
  {\doibase 10.1103/PhysRevLett.96.162302} {\bibfield  {journal} {\bibinfo
  {journal} {Phys. Rev. Lett.}\ }\textbf {\bibinfo {volume} {96}},\ \bibinfo
  {pages} {162302} (\bibinfo {year} {2006})},\ \Eprint
  {http://arxiv.org/abs/nucl-ex/0605007} {arXiv:nucl-ex/0605007} \BibitemShut
  {NoStop}%
\bibitem [{\citenamefont {Son}\ and\ \citenamefont
  {Stephanov}(2001)}]{Son:2000xc}%
  \BibitemOpen
  \bibfield  {author} {\bibinfo {author} {\bibfnamefont {D.~T.}\ \bibnamefont
  {Son}}\ and\ \bibinfo {author} {\bibfnamefont {M.~A.}\ \bibnamefont
  {Stephanov}},\ }\href {\doibase 10.1103/PhysRevLett.86.592} {\bibfield
  {journal} {\bibinfo  {journal} {Phys. Rev. Lett.}\ }\textbf {\bibinfo
  {volume} {86}},\ \bibinfo {pages} {592} (\bibinfo {year} {2001})},\ \Eprint
  {http://arxiv.org/abs/hep-ph/0005225} {arXiv:hep-ph/0005225} \BibitemShut
  {NoStop}%
\bibitem [{\citenamefont {Fukushima}\ and\ \citenamefont
  {Sasaki}(2013)}]{Fukushima:2013rx}%
  \BibitemOpen
  \bibfield  {author} {\bibinfo {author} {\bibfnamefont {K.}~\bibnamefont
  {Fukushima}}\ and\ \bibinfo {author} {\bibfnamefont {C.}~\bibnamefont
  {Sasaki}},\ }\href {\doibase 10.1016/j.ppnp.2013.05.003} {\bibfield
  {journal} {\bibinfo  {journal} {Prog. Part. Nucl. Phys.}\ }\textbf {\bibinfo
  {volume} {72}},\ \bibinfo {pages} {99} (\bibinfo {year} {2013})},\ \Eprint
  {http://arxiv.org/abs/1301.6377} {arXiv:1301.6377 [hep-ph]} \BibitemShut
  {NoStop}%
\bibitem [{\citenamefont {Aarts}\ \emph {et~al.}(2023)\citenamefont {Aarts}
  \emph {et~al.}}]{Aarts:2023vsf}%
  \BibitemOpen
  \bibfield  {author} {\bibinfo {author} {\bibfnamefont {G.}~\bibnamefont
  {Aarts}} \emph {et~al.},\ }\href {\doibase 10.1016/j.ppnp.2023.104070}
  {\bibfield  {journal} {\bibinfo  {journal} {Prog. Part. Nucl. Phys.}\
  }\textbf {\bibinfo {volume} {133}},\ \bibinfo {pages} {104070} (\bibinfo
  {year} {2023})},\ \Eprint {http://arxiv.org/abs/2301.04382} {arXiv:2301.04382
  [hep-lat]} \BibitemShut {NoStop}%
\bibitem [{\citenamefont {Kogut}\ and\ \citenamefont
  {Toublan}(2001)}]{Kogut:2001id}%
  \BibitemOpen
  \bibfield  {author} {\bibinfo {author} {\bibfnamefont {J.~B.}\ \bibnamefont
  {Kogut}}\ and\ \bibinfo {author} {\bibfnamefont {D.}~\bibnamefont
  {Toublan}},\ }\href {\doibase 10.1103/PhysRevD.64.034007} {\bibfield
  {journal} {\bibinfo  {journal} {Phys. Rev. D}\ }\textbf {\bibinfo {volume}
  {64}},\ \bibinfo {pages} {034007} (\bibinfo {year} {2001})},\ \Eprint
  {http://arxiv.org/abs/hep-ph/0103271} {arXiv:hep-ph/0103271} \BibitemShut
  {NoStop}%
\bibitem [{\citenamefont {Kogut}\ and\ \citenamefont
  {Sinclair}(2002)}]{Kogut:2002zg}%
  \BibitemOpen
  \bibfield  {author} {\bibinfo {author} {\bibfnamefont {J.~B.}\ \bibnamefont
  {Kogut}}\ and\ \bibinfo {author} {\bibfnamefont {D.~K.}\ \bibnamefont
  {Sinclair}},\ }\href {\doibase 10.1103/PhysRevD.66.034505} {\bibfield
  {journal} {\bibinfo  {journal} {Phys. Rev. D}\ }\textbf {\bibinfo {volume}
  {66}},\ \bibinfo {pages} {034505} (\bibinfo {year} {2002})},\ \Eprint
  {http://arxiv.org/abs/hep-lat/0202028} {arXiv:hep-lat/0202028} \BibitemShut
  {NoStop}%
\bibitem [{\citenamefont {Brandt}\ \emph
  {et~al.}(2018{\natexlab{a}})\citenamefont {Brandt}, \citenamefont {Endrodi},\
  and\ \citenamefont {Schmalzbauer}}]{Brandt:2017oyy}%
  \BibitemOpen
  \bibfield  {author} {\bibinfo {author} {\bibfnamefont {B.~B.}\ \bibnamefont
  {Brandt}}, \bibinfo {author} {\bibfnamefont {G.}~\bibnamefont {Endrodi}}, \
  and\ \bibinfo {author} {\bibfnamefont {S.}~\bibnamefont {Schmalzbauer}},\
  }\href {\doibase 10.1103/PhysRevD.97.054514} {\bibfield  {journal} {\bibinfo
  {journal} {Phys. Rev. D}\ }\textbf {\bibinfo {volume} {97}},\ \bibinfo
  {pages} {054514} (\bibinfo {year} {2018}{\natexlab{a}})},\ \Eprint
  {http://arxiv.org/abs/1712.08190} {arXiv:1712.08190 [hep-lat]} \BibitemShut
  {NoStop}%
\bibitem [{\citenamefont {Brandt}\ \emph
  {et~al.}(2018{\natexlab{b}})\citenamefont {Brandt}, \citenamefont {Endrodi},
  \citenamefont {Fraga}, \citenamefont {Hippert}, \citenamefont
  {Schaffner-Bielich},\ and\ \citenamefont {Schmalzbauer}}]{Brandt:2018bwq}%
  \BibitemOpen
  \bibfield  {author} {\bibinfo {author} {\bibfnamefont {B.~B.}\ \bibnamefont
  {Brandt}}, \bibinfo {author} {\bibfnamefont {G.}~\bibnamefont {Endrodi}},
  \bibinfo {author} {\bibfnamefont {E.~S.}\ \bibnamefont {Fraga}}, \bibinfo
  {author} {\bibfnamefont {M.}~\bibnamefont {Hippert}}, \bibinfo {author}
  {\bibfnamefont {J.}~\bibnamefont {Schaffner-Bielich}}, \ and\ \bibinfo
  {author} {\bibfnamefont {S.}~\bibnamefont {Schmalzbauer}},\ }\href {\doibase
  10.1103/PhysRevD.98.094510} {\bibfield  {journal} {\bibinfo  {journal} {Phys.
  Rev. D}\ }\textbf {\bibinfo {volume} {98}},\ \bibinfo {pages} {094510}
  (\bibinfo {year} {2018}{\natexlab{b}})},\ \Eprint
  {http://arxiv.org/abs/1802.06685} {arXiv:1802.06685 [hep-ph]} \BibitemShut
  {NoStop}%
\bibitem [{\citenamefont {Nambu}\ and\ \citenamefont
  {Jona-Lasinio}(1961{\natexlab{a}})}]{Nambu:1961tp}%
  \BibitemOpen
  \bibfield  {author} {\bibinfo {author} {\bibfnamefont {Y.}~\bibnamefont
  {Nambu}}\ and\ \bibinfo {author} {\bibfnamefont {G.}~\bibnamefont
  {Jona-Lasinio}},\ }\href {\doibase 10.1103/PhysRev.122.345} {\bibfield
  {journal} {\bibinfo  {journal} {Phys. Rev.}\ }\textbf {\bibinfo {volume}
  {122}},\ \bibinfo {pages} {345} (\bibinfo {year}
  {1961}{\natexlab{a}})}\BibitemShut {NoStop}%
\bibitem [{\citenamefont {Nambu}\ and\ \citenamefont
  {Jona-Lasinio}(1961{\natexlab{b}})}]{Nambu:1961fr}%
  \BibitemOpen
  \bibfield  {author} {\bibinfo {author} {\bibfnamefont {Y.}~\bibnamefont
  {Nambu}}\ and\ \bibinfo {author} {\bibfnamefont {G.}~\bibnamefont
  {Jona-Lasinio}},\ }\href {\doibase 10.1103/PhysRev.124.246} {\bibfield
  {journal} {\bibinfo  {journal} {Phys. Rev.}\ }\textbf {\bibinfo {volume}
  {124}},\ \bibinfo {pages} {246} (\bibinfo {year}
  {1961}{\natexlab{b}})}\BibitemShut {NoStop}%
\bibitem [{\citenamefont {Turko}(1994)}]{Turko:1993dy}%
  \BibitemOpen
  \bibfield  {author} {\bibinfo {author} {\bibfnamefont {L.}~\bibnamefont
  {Turko}},\ }\href {\doibase 10.1007/BF01413108} {\bibfield  {journal}
  {\bibinfo  {journal} {Z. Phys. C}\ }\textbf {\bibinfo {volume} {61}},\
  \bibinfo {pages} {297} (\bibinfo {year} {1994})}\BibitemShut {NoStop}%
\bibitem [{\citenamefont {Abuki}\ \emph {et~al.}(2009)\citenamefont {Abuki},
  \citenamefont {Anglani}, \citenamefont {Gatto}, \citenamefont {Pellicoro},\
  and\ \citenamefont {Ruggieri}}]{Abuki:2008wm}%
  \BibitemOpen
  \bibfield  {author} {\bibinfo {author} {\bibfnamefont {H.}~\bibnamefont
  {Abuki}}, \bibinfo {author} {\bibfnamefont {R.}~\bibnamefont {Anglani}},
  \bibinfo {author} {\bibfnamefont {R.}~\bibnamefont {Gatto}}, \bibinfo
  {author} {\bibfnamefont {M.}~\bibnamefont {Pellicoro}}, \ and\ \bibinfo
  {author} {\bibfnamefont {M.}~\bibnamefont {Ruggieri}},\ }\href {\doibase
  10.1103/PhysRevD.79.034032} {\bibfield  {journal} {\bibinfo  {journal} {Phys.
  Rev. D}\ }\textbf {\bibinfo {volume} {79}},\ \bibinfo {pages} {034032}
  (\bibinfo {year} {2009})},\ \Eprint {http://arxiv.org/abs/0809.2658}
  {arXiv:0809.2658 [hep-ph]} \BibitemShut {NoStop}%
\bibitem [{\citenamefont {Andersen}\ and\ \citenamefont
  {Kyllingstad}(2009)}]{Andersen:2007qv}%
  \BibitemOpen
  \bibfield  {author} {\bibinfo {author} {\bibfnamefont {J.~O.}\ \bibnamefont
  {Andersen}}\ and\ \bibinfo {author} {\bibfnamefont {L.}~\bibnamefont
  {Kyllingstad}},\ }\href {\doibase 10.1088/0954-3899/37/1/015003} {\bibfield
  {journal} {\bibinfo  {journal} {J. Phys. G}\ }\textbf {\bibinfo {volume}
  {37}},\ \bibinfo {pages} {015003} (\bibinfo {year} {2009})},\ \Eprint
  {http://arxiv.org/abs/hep-ph/0701033} {arXiv:hep-ph/0701033} \BibitemShut
  {NoStop}%
\bibitem [{\citenamefont {Sun}\ \emph {et~al.}(2007)\citenamefont {Sun},
  \citenamefont {He},\ and\ \citenamefont {Zhuang}}]{Sun:2007fc}%
  \BibitemOpen
  \bibfield  {author} {\bibinfo {author} {\bibfnamefont {G.-f.}\ \bibnamefont
  {Sun}}, \bibinfo {author} {\bibfnamefont {L.}~\bibnamefont {He}}, \ and\
  \bibinfo {author} {\bibfnamefont {P.}~\bibnamefont {Zhuang}},\ }\href
  {\doibase 10.1103/PhysRevD.75.096004} {\bibfield  {journal} {\bibinfo
  {journal} {Phys. Rev. D}\ }\textbf {\bibinfo {volume} {75}},\ \bibinfo
  {pages} {096004} (\bibinfo {year} {2007})},\ \Eprint
  {http://arxiv.org/abs/hep-ph/0703159} {arXiv:hep-ph/0703159} \BibitemShut
  {NoStop}%
\bibitem [{\citenamefont {Ebert}\ and\ \citenamefont
  {Klimenko}(2006{\natexlab{a}})}]{Ebert:2005wr}%
  \BibitemOpen
  \bibfield  {author} {\bibinfo {author} {\bibfnamefont {D.}~\bibnamefont
  {Ebert}}\ and\ \bibinfo {author} {\bibfnamefont {K.~G.}\ \bibnamefont
  {Klimenko}},\ }\href {\doibase 10.1140/epjc/s2006-02527-5} {\bibfield
  {journal} {\bibinfo  {journal} {Eur. Phys. J. C}\ }\textbf {\bibinfo {volume}
  {46}},\ \bibinfo {pages} {771} (\bibinfo {year} {2006}{\natexlab{a}})},\
  \Eprint {http://arxiv.org/abs/hep-ph/0510222} {arXiv:hep-ph/0510222}
  \BibitemShut {NoStop}%
\bibitem [{\citenamefont {Ebert}\ and\ \citenamefont
  {Klimenko}(2006{\natexlab{b}})}]{Ebert:2005cs}%
  \BibitemOpen
  \bibfield  {author} {\bibinfo {author} {\bibfnamefont {D.}~\bibnamefont
  {Ebert}}\ and\ \bibinfo {author} {\bibfnamefont {K.~G.}\ \bibnamefont
  {Klimenko}},\ }\href {\doibase 10.1088/0954-3899/32/5/001} {\bibfield
  {journal} {\bibinfo  {journal} {J. Phys. G}\ }\textbf {\bibinfo {volume}
  {32}},\ \bibinfo {pages} {599} (\bibinfo {year} {2006}{\natexlab{b}})},\
  \Eprint {http://arxiv.org/abs/hep-ph/0507007} {arXiv:hep-ph/0507007}
  \BibitemShut {NoStop}%
\bibitem [{\citenamefont {He}\ \emph {et~al.}(2006)\citenamefont {He},
  \citenamefont {Jin},\ and\ \citenamefont {Zhuang}}]{He:2006tn}%
  \BibitemOpen
  \bibfield  {author} {\bibinfo {author} {\bibfnamefont {L.}~\bibnamefont
  {He}}, \bibinfo {author} {\bibfnamefont {M.}~\bibnamefont {Jin}}, \ and\
  \bibinfo {author} {\bibfnamefont {P.}~\bibnamefont {Zhuang}},\ }\href
  {\doibase 10.1103/PhysRevD.74.036005} {\bibfield  {journal} {\bibinfo
  {journal} {Phys. Rev. D}\ }\textbf {\bibinfo {volume} {74}},\ \bibinfo
  {pages} {036005} (\bibinfo {year} {2006})},\ \Eprint
  {http://arxiv.org/abs/hep-ph/0604224} {arXiv:hep-ph/0604224} \BibitemShut
  {NoStop}%
\bibitem [{\citenamefont {He}\ \emph {et~al.}(2005)\citenamefont {He},
  \citenamefont {Jin},\ and\ \citenamefont {Zhuang}}]{He:2005nk}%
  \BibitemOpen
  \bibfield  {author} {\bibinfo {author} {\bibfnamefont {L.-y.}\ \bibnamefont
  {He}}, \bibinfo {author} {\bibfnamefont {M.}~\bibnamefont {Jin}}, \ and\
  \bibinfo {author} {\bibfnamefont {P.-f.}\ \bibnamefont {Zhuang}},\ }\href
  {\doibase 10.1103/PhysRevD.71.116001} {\bibfield  {journal} {\bibinfo
  {journal} {Phys. Rev. D}\ }\textbf {\bibinfo {volume} {71}},\ \bibinfo
  {pages} {116001} (\bibinfo {year} {2005})},\ \Eprint
  {http://arxiv.org/abs/hep-ph/0503272} {arXiv:hep-ph/0503272} \BibitemShut
  {NoStop}%
\bibitem [{\citenamefont {He}\ and\ \citenamefont {Zhuang}(2005)}]{He:2005sp}%
  \BibitemOpen
  \bibfield  {author} {\bibinfo {author} {\bibfnamefont {L.}~\bibnamefont
  {He}}\ and\ \bibinfo {author} {\bibfnamefont {P.}~\bibnamefont {Zhuang}},\
  }\href {\doibase 10.1016/j.physletb.2005.03.066} {\bibfield  {journal}
  {\bibinfo  {journal} {Phys. Lett. B}\ }\textbf {\bibinfo {volume} {615}},\
  \bibinfo {pages} {93} (\bibinfo {year} {2005})},\ \Eprint
  {http://arxiv.org/abs/hep-ph/0501024} {arXiv:hep-ph/0501024} \BibitemShut
  {NoStop}%
\bibitem [{\citenamefont {Barducci}\ \emph {et~al.}(2004)\citenamefont
  {Barducci}, \citenamefont {Casalbuoni}, \citenamefont {Pettini},\ and\
  \citenamefont {Ravagli}}]{Barducci:2004tt}%
  \BibitemOpen
  \bibfield  {author} {\bibinfo {author} {\bibfnamefont {A.}~\bibnamefont
  {Barducci}}, \bibinfo {author} {\bibfnamefont {R.}~\bibnamefont
  {Casalbuoni}}, \bibinfo {author} {\bibfnamefont {G.}~\bibnamefont {Pettini}},
  \ and\ \bibinfo {author} {\bibfnamefont {L.}~\bibnamefont {Ravagli}},\ }\href
  {\doibase 10.1103/PhysRevD.69.096004} {\bibfield  {journal} {\bibinfo
  {journal} {Phys. Rev. D}\ }\textbf {\bibinfo {volume} {69}},\ \bibinfo
  {pages} {096004} (\bibinfo {year} {2004})},\ \Eprint
  {http://arxiv.org/abs/hep-ph/0402104} {arXiv:hep-ph/0402104} \BibitemShut
  {NoStop}%
\bibitem [{\citenamefont {Toublan}\ and\ \citenamefont
  {Kogut}(2003)}]{Toublan:2003tt}%
  \BibitemOpen
  \bibfield  {author} {\bibinfo {author} {\bibfnamefont {D.}~\bibnamefont
  {Toublan}}\ and\ \bibinfo {author} {\bibfnamefont {J.~B.}\ \bibnamefont
  {Kogut}},\ }\href {\doibase 10.1016/S0370-2693(03)00701-9} {\bibfield
  {journal} {\bibinfo  {journal} {Phys. Lett. B}\ }\textbf {\bibinfo {volume}
  {564}},\ \bibinfo {pages} {212} (\bibinfo {year} {2003})},\ \Eprint
  {http://arxiv.org/abs/hep-ph/0301183} {arXiv:hep-ph/0301183} \BibitemShut
  {NoStop}%
\bibitem [{\citenamefont {Frank}\ \emph {et~al.}(2003)\citenamefont {Frank},
  \citenamefont {Buballa},\ and\ \citenamefont {Oertel}}]{Frank:2003ve}%
  \BibitemOpen
  \bibfield  {author} {\bibinfo {author} {\bibfnamefont {M.}~\bibnamefont
  {Frank}}, \bibinfo {author} {\bibfnamefont {M.}~\bibnamefont {Buballa}}, \
  and\ \bibinfo {author} {\bibfnamefont {M.}~\bibnamefont {Oertel}},\ }\href
  {\doibase 10.1016/S0370-2693(03)00607-5} {\bibfield  {journal} {\bibinfo
  {journal} {Phys. Lett. B}\ }\textbf {\bibinfo {volume} {562}},\ \bibinfo
  {pages} {221} (\bibinfo {year} {2003})},\ \Eprint
  {http://arxiv.org/abs/hep-ph/0303109} {arXiv:hep-ph/0303109} \BibitemShut
  {NoStop}%
\bibitem [{\citenamefont {Mu}\ \emph {et~al.}(2010)\citenamefont {Mu},
  \citenamefont {He},\ and\ \citenamefont {Liu}}]{Mu:2010zz}%
  \BibitemOpen
  \bibfield  {author} {\bibinfo {author} {\bibfnamefont {C.-f.}\ \bibnamefont
  {Mu}}, \bibinfo {author} {\bibfnamefont {L.-y.}\ \bibnamefont {He}}, \ and\
  \bibinfo {author} {\bibfnamefont {Y.-x.}\ \bibnamefont {Liu}},\ }\href
  {\doibase 10.1103/PhysRevD.82.056006} {\bibfield  {journal} {\bibinfo
  {journal} {Phys. Rev. D}\ }\textbf {\bibinfo {volume} {82}},\ \bibinfo
  {pages} {056006} (\bibinfo {year} {2010})}\BibitemShut {NoStop}%
\bibitem [{\citenamefont {Xia}\ \emph {et~al.}(2013)\citenamefont {Xia},
  \citenamefont {He},\ and\ \citenamefont {Zhuang}}]{Xia:2013caa}%
  \BibitemOpen
  \bibfield  {author} {\bibinfo {author} {\bibfnamefont {T.}~\bibnamefont
  {Xia}}, \bibinfo {author} {\bibfnamefont {L.}~\bibnamefont {He}}, \ and\
  \bibinfo {author} {\bibfnamefont {P.}~\bibnamefont {Zhuang}},\ }\href
  {\doibase 10.1103/PhysRevD.88.056013} {\bibfield  {journal} {\bibinfo
  {journal} {Phys. Rev. D}\ }\textbf {\bibinfo {volume} {88}},\ \bibinfo
  {pages} {056013} (\bibinfo {year} {2013})},\ \Eprint
  {http://arxiv.org/abs/1307.4622} {arXiv:1307.4622 [hep-ph]} \BibitemShut
  {NoStop}%
\bibitem [{\citenamefont {Ebert}\ \emph {et~al.}(2016)\citenamefont {Ebert},
  \citenamefont {Khunjua},\ and\ \citenamefont {Klimenko}}]{Ebert:2016hkd}%
  \BibitemOpen
  \bibfield  {author} {\bibinfo {author} {\bibfnamefont {D.}~\bibnamefont
  {Ebert}}, \bibinfo {author} {\bibfnamefont {T.~G.}\ \bibnamefont {Khunjua}},
  \ and\ \bibinfo {author} {\bibfnamefont {K.~G.}\ \bibnamefont {Klimenko}},\
  }\href {\doibase 10.1103/PhysRevD.94.116016} {\bibfield  {journal} {\bibinfo
  {journal} {Phys. Rev. D}\ }\textbf {\bibinfo {volume} {94}},\ \bibinfo
  {pages} {116016} (\bibinfo {year} {2016})},\ \Eprint
  {http://arxiv.org/abs/1608.07688} {arXiv:1608.07688 [hep-ph]} \BibitemShut
  {NoStop}%
\bibitem [{\citenamefont {Khunjua}\ \emph {et~al.}(2017)\citenamefont
  {Khunjua}, \citenamefont {Klimenko}, \citenamefont {Zhokhov},\ and\
  \citenamefont {Zhukovsky}}]{Khunjua:2017khh}%
  \BibitemOpen
  \bibfield  {author} {\bibinfo {author} {\bibfnamefont {T.~G.}\ \bibnamefont
  {Khunjua}}, \bibinfo {author} {\bibfnamefont {K.~G.}\ \bibnamefont
  {Klimenko}}, \bibinfo {author} {\bibfnamefont {R.~N.}\ \bibnamefont
  {Zhokhov}}, \ and\ \bibinfo {author} {\bibfnamefont {V.~C.}\ \bibnamefont
  {Zhukovsky}},\ }\href {\doibase 10.1103/PhysRevD.95.105010} {\bibfield
  {journal} {\bibinfo  {journal} {Phys. Rev. D}\ }\textbf {\bibinfo {volume}
  {95}},\ \bibinfo {pages} {105010} (\bibinfo {year} {2017})},\ \Eprint
  {http://arxiv.org/abs/1704.01477} {arXiv:1704.01477 [hep-ph]} \BibitemShut
  {NoStop}%
\bibitem [{\citenamefont {Khunjua}\ \emph
  {et~al.}(2019{\natexlab{a}})\citenamefont {Khunjua}, \citenamefont
  {Klimenko},\ and\ \citenamefont {Zhokhov}}]{Khunjua:2019lbv}%
  \BibitemOpen
  \bibfield  {author} {\bibinfo {author} {\bibfnamefont {T.~G.}\ \bibnamefont
  {Khunjua}}, \bibinfo {author} {\bibfnamefont {K.~G.}\ \bibnamefont
  {Klimenko}}, \ and\ \bibinfo {author} {\bibfnamefont {R.~N.}\ \bibnamefont
  {Zhokhov}},\ }\href {\doibase 10.1007/JHEP06(2019)006} {\bibfield  {journal}
  {\bibinfo  {journal} {JHEP}\ }\textbf {\bibinfo {volume} {06}},\ \bibinfo
  {pages} {006} (\bibinfo {year} {2019}{\natexlab{a}})},\ \Eprint
  {http://arxiv.org/abs/1901.02855} {arXiv:1901.02855 [hep-ph]} \BibitemShut
  {NoStop}%
\bibitem [{\citenamefont {Khunjua}\ \emph
  {et~al.}(2019{\natexlab{b}})\citenamefont {Khunjua}, \citenamefont
  {Klimenko},\ and\ \citenamefont {Zhokhov}}]{Khunjua:2019ini}%
  \BibitemOpen
  \bibfield  {author} {\bibinfo {author} {\bibfnamefont {T.~G.}\ \bibnamefont
  {Khunjua}}, \bibinfo {author} {\bibfnamefont {K.~G.}\ \bibnamefont
  {Klimenko}}, \ and\ \bibinfo {author} {\bibfnamefont {R.~N.}\ \bibnamefont
  {Zhokhov}},\ }\href {\doibase 10.1103/PhysRevD.100.034009} {\bibfield
  {journal} {\bibinfo  {journal} {Phys. Rev. D}\ }\textbf {\bibinfo {volume}
  {100}},\ \bibinfo {pages} {034009} (\bibinfo {year} {2019}{\natexlab{b}})},\
  \Eprint {http://arxiv.org/abs/1907.04151} {arXiv:1907.04151 [hep-ph]}
  \BibitemShut {NoStop}%
\bibitem [{\citenamefont {Khunjua}\ \emph {et~al.}(2020)\citenamefont
  {Khunjua}, \citenamefont {Klimenko},\ and\ \citenamefont
  {Zhokhov}}]{Khunjua:2020xws}%
  \BibitemOpen
  \bibfield  {author} {\bibinfo {author} {\bibfnamefont {T.~G.}\ \bibnamefont
  {Khunjua}}, \bibinfo {author} {\bibfnamefont {K.~G.}\ \bibnamefont
  {Klimenko}}, \ and\ \bibinfo {author} {\bibfnamefont {R.~N.}\ \bibnamefont
  {Zhokhov}},\ }\href {\doibase 10.1007/JHEP06(2020)148} {\bibfield  {journal}
  {\bibinfo  {journal} {JHEP}\ }\textbf {\bibinfo {volume} {06}},\ \bibinfo
  {pages} {148} (\bibinfo {year} {2020})},\ \Eprint
  {http://arxiv.org/abs/2003.10562} {arXiv:2003.10562 [hep-ph]} \BibitemShut
  {NoStop}%
\bibitem [{\citenamefont {Lu}\ \emph {et~al.}(2020)\citenamefont {Lu},
  \citenamefont {Xia},\ and\ \citenamefont {Ruggieri}}]{Lu:2019diy}%
  \BibitemOpen
  \bibfield  {author} {\bibinfo {author} {\bibfnamefont {Z.-Y.}\ \bibnamefont
  {Lu}}, \bibinfo {author} {\bibfnamefont {C.-J.}\ \bibnamefont {Xia}}, \ and\
  \bibinfo {author} {\bibfnamefont {M.}~\bibnamefont {Ruggieri}},\ }\href
  {\doibase 10.1140/epjc/s10052-020-7614-6} {\bibfield  {journal} {\bibinfo
  {journal} {Eur. Phys. J. C}\ }\textbf {\bibinfo {volume} {80}},\ \bibinfo
  {pages} {46} (\bibinfo {year} {2020})},\ \Eprint
  {http://arxiv.org/abs/1907.11497} {arXiv:1907.11497 [hep-ph]} \BibitemShut
  {NoStop}%
\bibitem [{\citenamefont {Avancini}\ \emph {et~al.}(2019)\citenamefont
  {Avancini}, \citenamefont {Bandyopadhyay}, \citenamefont {Duarte},\ and\
  \citenamefont {Farias}}]{Avancini:2019ego}%
  \BibitemOpen
  \bibfield  {author} {\bibinfo {author} {\bibfnamefont {S.~S.}\ \bibnamefont
  {Avancini}}, \bibinfo {author} {\bibfnamefont {A.}~\bibnamefont
  {Bandyopadhyay}}, \bibinfo {author} {\bibfnamefont {D.~C.}\ \bibnamefont
  {Duarte}}, \ and\ \bibinfo {author} {\bibfnamefont {R.~L.~S.}\ \bibnamefont
  {Farias}},\ }\href {\doibase 10.1103/PhysRevD.100.116002} {\bibfield
  {journal} {\bibinfo  {journal} {Phys. Rev. D}\ }\textbf {\bibinfo {volume}
  {100}},\ \bibinfo {pages} {116002} (\bibinfo {year} {2019})},\ \Eprint
  {http://arxiv.org/abs/1907.09880} {arXiv:1907.09880 [hep-ph]} \BibitemShut
  {NoStop}%
\bibitem [{\citenamefont {Lopes}\ \emph {et~al.}(2021)\citenamefont {Lopes},
  \citenamefont {Avancini}, \citenamefont {Bandyopadhyay}, \citenamefont
  {Duarte},\ and\ \citenamefont {Farias}}]{Lopes:2021tro}%
  \BibitemOpen
  \bibfield  {author} {\bibinfo {author} {\bibfnamefont {B.~S.}\ \bibnamefont
  {Lopes}}, \bibinfo {author} {\bibfnamefont {S.~S.}\ \bibnamefont {Avancini}},
  \bibinfo {author} {\bibfnamefont {A.}~\bibnamefont {Bandyopadhyay}}, \bibinfo
  {author} {\bibfnamefont {D.~C.}\ \bibnamefont {Duarte}}, \ and\ \bibinfo
  {author} {\bibfnamefont {R.~L.~S.}\ \bibnamefont {Farias}},\ }\href {\doibase
  10.1103/PhysRevD.103.076023} {\bibfield  {journal} {\bibinfo  {journal}
  {Phys. Rev. D}\ }\textbf {\bibinfo {volume} {103}},\ \bibinfo {pages}
  {076023} (\bibinfo {year} {2021})},\ \Eprint
  {http://arxiv.org/abs/2102.02844} {arXiv:2102.02844 [hep-ph]} \BibitemShut
  {NoStop}%
\bibitem [{\citenamefont {Klevansky}(1992)}]{Klevansky:1992qe}%
  \BibitemOpen
  \bibfield  {author} {\bibinfo {author} {\bibfnamefont {S.~P.}\ \bibnamefont
  {Klevansky}},\ }\href {\doibase 10.1103/RevModPhys.64.649} {\bibfield
  {journal} {\bibinfo  {journal} {Rev. Mod. Phys.}\ }\textbf {\bibinfo {volume}
  {64}},\ \bibinfo {pages} {649} (\bibinfo {year} {1992})}\BibitemShut
  {NoStop}%
\bibitem [{\citenamefont {Buballa}(2005)}]{Buballa:2003qv}%
  \BibitemOpen
  \bibfield  {author} {\bibinfo {author} {\bibfnamefont {M.}~\bibnamefont
  {Buballa}},\ }\href {\doibase 10.1016/j.physrep.2004.11.004} {\bibfield
  {journal} {\bibinfo  {journal} {Phys. Rept.}\ }\textbf {\bibinfo {volume}
  {407}},\ \bibinfo {pages} {205} (\bibinfo {year} {2005})},\ \Eprint
  {http://arxiv.org/abs/hep-ph/0402234} {arXiv:hep-ph/0402234} \BibitemShut
  {NoStop}%
\bibitem [{\citenamefont {Fukushima}(2008)}]{Fukushima:2008wg}%
  \BibitemOpen
  \bibfield  {author} {\bibinfo {author} {\bibfnamefont {K.}~\bibnamefont
  {Fukushima}},\ }\href {\doibase 10.1103/PhysRevD.77.114028} {\bibfield
  {journal} {\bibinfo  {journal} {Phys. Rev. D}\ }\textbf {\bibinfo {volume}
  {77}},\ \bibinfo {pages} {114028} (\bibinfo {year} {2008})},\ \bibinfo {note}
  {[Erratum: Phys.Rev.D 78, 039902 (2008)]},\ \Eprint
  {http://arxiv.org/abs/0803.3318} {arXiv:0803.3318 [hep-ph]} \BibitemShut
  {NoStop}%
\bibitem [{\citenamefont {Kojo}\ \emph {et~al.}(2015)\citenamefont {Kojo},
  \citenamefont {Powell}, \citenamefont {Song},\ and\ \citenamefont
  {Baym}}]{Kojo:2014rca}%
  \BibitemOpen
  \bibfield  {author} {\bibinfo {author} {\bibfnamefont {T.}~\bibnamefont
  {Kojo}}, \bibinfo {author} {\bibfnamefont {P.~D.}\ \bibnamefont {Powell}},
  \bibinfo {author} {\bibfnamefont {Y.}~\bibnamefont {Song}}, \ and\ \bibinfo
  {author} {\bibfnamefont {G.}~\bibnamefont {Baym}},\ }\href {\doibase
  10.1103/PhysRevD.91.045003} {\bibfield  {journal} {\bibinfo  {journal} {Phys.
  Rev. D}\ }\textbf {\bibinfo {volume} {91}},\ \bibinfo {pages} {045003}
  (\bibinfo {year} {2015})},\ \Eprint {http://arxiv.org/abs/1412.1108}
  {arXiv:1412.1108 [hep-ph]} \BibitemShut {NoStop}%
\bibitem [{\citenamefont {Ali}\ \emph {et~al.}(2025)\citenamefont {Ali},
  \citenamefont {Biswas},\ and\ \citenamefont {Islam}}]{Ali:2024owl}%
  \BibitemOpen
  \bibfield  {author} {\bibinfo {author} {\bibfnamefont {M.~S.}\ \bibnamefont
  {Ali}}, \bibinfo {author} {\bibfnamefont {D.}~\bibnamefont {Biswas}}, \ and\
  \bibinfo {author} {\bibfnamefont {C.~A.}\ \bibnamefont {Islam}},\ }\href
  {\doibase 10.1140/epja/s10050-025-01718-y} {\bibfield  {journal} {\bibinfo
  {journal} {Eur. Phys. J. A}\ }\textbf {\bibinfo {volume} {61}},\ \bibinfo
  {pages} {240} (\bibinfo {year} {2025})},\ \Eprint
  {http://arxiv.org/abs/2410.00653} {arXiv:2410.00653 [nucl-th]} \BibitemShut
  {NoStop}%
\bibitem [{\citenamefont {Sasaki}\ \emph {et~al.}(2007)\citenamefont {Sasaki},
  \citenamefont {Friman},\ and\ \citenamefont {Redlich}}]{Sasaki:2006ws}%
  \BibitemOpen
  \bibfield  {author} {\bibinfo {author} {\bibfnamefont {C.}~\bibnamefont
  {Sasaki}}, \bibinfo {author} {\bibfnamefont {B.}~\bibnamefont {Friman}}, \
  and\ \bibinfo {author} {\bibfnamefont {K.}~\bibnamefont {Redlich}},\ }\href
  {\doibase 10.1103/PhysRevD.75.054026} {\bibfield  {journal} {\bibinfo
  {journal} {Phys. Rev. D}\ }\textbf {\bibinfo {volume} {75}},\ \bibinfo
  {pages} {054026} (\bibinfo {year} {2007})},\ \Eprint
  {http://arxiv.org/abs/hep-ph/0611143} {arXiv:hep-ph/0611143} \BibitemShut
  {NoStop}%
\bibitem [{\citenamefont {Davidson}\ and\ \citenamefont
  {Ruiz~Arriola}(1995)}]{Davidson:1995fq}%
  \BibitemOpen
  \bibfield  {author} {\bibinfo {author} {\bibfnamefont {R.~M.}\ \bibnamefont
  {Davidson}}\ and\ \bibinfo {author} {\bibfnamefont {E.}~\bibnamefont
  {Ruiz~Arriola}},\ }\href {\doibase 10.1016/0370-2693(95)01119-B} {\bibfield
  {journal} {\bibinfo  {journal} {Phys. Lett. B}\ }\textbf {\bibinfo {volume}
  {359}},\ \bibinfo {pages} {273} (\bibinfo {year} {1995})}\BibitemShut
  {NoStop}%
\bibitem [{\citenamefont {Islam}\ \emph {et~al.}(2015)\citenamefont {Islam},
  \citenamefont {Majumder}, \citenamefont {Haque},\ and\ \citenamefont
  {Mustafa}}]{Islam:2014sea}%
  \BibitemOpen
  \bibfield  {author} {\bibinfo {author} {\bibfnamefont {C.~A.}\ \bibnamefont
  {Islam}}, \bibinfo {author} {\bibfnamefont {S.}~\bibnamefont {Majumder}},
  \bibinfo {author} {\bibfnamefont {N.}~\bibnamefont {Haque}}, \ and\ \bibinfo
  {author} {\bibfnamefont {M.~G.}\ \bibnamefont {Mustafa}},\ }\href {\doibase
  10.1007/JHEP02(2015)011} {\bibfield  {journal} {\bibinfo  {journal} {JHEP}\
  }\textbf {\bibinfo {volume} {02}},\ \bibinfo {pages} {011} (\bibinfo {year}
  {2015})},\ \Eprint {http://arxiv.org/abs/1411.6407} {arXiv:1411.6407
  [hep-ph]} \BibitemShut {NoStop}%
\bibitem [{\citenamefont {Ablyazimov}\ \emph {et~al.}(2017)\citenamefont
  {Ablyazimov} \emph {et~al.}}]{CBM:2016kpk}%
  \BibitemOpen
  \bibfield  {author} {\bibinfo {author} {\bibfnamefont {T.}~\bibnamefont
  {Ablyazimov}} \emph {et~al.} (\bibinfo {collaboration} {CBM}),\ }\href
  {\doibase 10.1140/epja/i2017-12248-y} {\bibfield  {journal} {\bibinfo
  {journal} {Eur. Phys. J. A}\ }\textbf {\bibinfo {volume} {53}},\ \bibinfo
  {pages} {60} (\bibinfo {year} {2017})},\ \Eprint
  {http://arxiv.org/abs/1607.01487} {arXiv:1607.01487 [nucl-ex]} \BibitemShut
  {NoStop}%
\bibitem [{\citenamefont {Sako}\ \emph {et~al.}(2014)\citenamefont {Sako} \emph
  {et~al.}}]{Sako:2014fha}%
  \BibitemOpen
  \bibfield  {author} {\bibinfo {author} {\bibfnamefont {H.}~\bibnamefont
  {Sako}} \emph {et~al.},\ }\href {\doibase 10.1016/j.nuclphysa.2014.08.065}
  {\bibfield  {journal} {\bibinfo  {journal} {Nucl. Phys. A}\ }\textbf
  {\bibinfo {volume} {931}},\ \bibinfo {pages} {1158} (\bibinfo {year}
  {2014})}\BibitemShut {NoStop}%
\bibitem [{\citenamefont {Kekelidze}\ \emph {et~al.}(2017)\citenamefont
  {Kekelidze}, \citenamefont {Kovalenko}, \citenamefont {Lednicky},
  \citenamefont {Matveev}, \citenamefont {Meshkov}, \citenamefont {Sorin},\
  and\ \citenamefont {Trubnikov}}]{Kekelidze:2017tgp}%
  \BibitemOpen
  \bibfield  {author} {\bibinfo {author} {\bibfnamefont {V.}~\bibnamefont
  {Kekelidze}}, \bibinfo {author} {\bibfnamefont {A.}~\bibnamefont
  {Kovalenko}}, \bibinfo {author} {\bibfnamefont {R.}~\bibnamefont {Lednicky}},
  \bibinfo {author} {\bibfnamefont {V.}~\bibnamefont {Matveev}}, \bibinfo
  {author} {\bibfnamefont {I.}~\bibnamefont {Meshkov}}, \bibinfo {author}
  {\bibfnamefont {A.}~\bibnamefont {Sorin}}, \ and\ \bibinfo {author}
  {\bibfnamefont {G.}~\bibnamefont {Trubnikov}},\ }\href {\doibase
  10.1016/j.nuclphysa.2017.06.031} {\bibfield  {journal} {\bibinfo  {journal}
  {Nucl. Phys. A}\ }\textbf {\bibinfo {volume} {967}},\ \bibinfo {pages} {884}
  (\bibinfo {year} {2017})}\BibitemShut {NoStop}%
\bibitem [{\citenamefont {Karsch}\ \emph {et~al.}(2001)\citenamefont {Karsch},
  \citenamefont {Mustafa},\ and\ \citenamefont {Thoma}}]{Karsch:2000gi}%
  \BibitemOpen
  \bibfield  {author} {\bibinfo {author} {\bibfnamefont {F.}~\bibnamefont
  {Karsch}}, \bibinfo {author} {\bibfnamefont {M.}~\bibnamefont {Mustafa}}, \
  and\ \bibinfo {author} {\bibfnamefont {M.}~\bibnamefont {Thoma}},\ }\href
  {\doibase 10.1016/S0370-2693(00)01322-8} {\bibfield  {journal} {\bibinfo
  {journal} {Phys. Lett.}\ }\textbf {\bibinfo {volume} {B497}},\ \bibinfo
  {pages} {249} (\bibinfo {year} {2001})},\ \Eprint
  {http://arxiv.org/abs/hep-ph/0007093} {arXiv:hep-ph/0007093 [hep-ph]}
  \BibitemShut {NoStop}%
\bibitem [{\citenamefont {Mustafa}\ \emph {et~al.}(2026)\citenamefont
  {Mustafa}, \citenamefont {Bandyopadhyay},\ and\ \citenamefont
  {Islam}}]{Mustafa:2025uad}%
  \BibitemOpen
  \bibfield  {author} {\bibinfo {author} {\bibfnamefont {M.~G.}\ \bibnamefont
  {Mustafa}}, \bibinfo {author} {\bibfnamefont {A.}~\bibnamefont
  {Bandyopadhyay}}, \ and\ \bibinfo {author} {\bibfnamefont {C.~A.}\
  \bibnamefont {Islam}},\ }\href {\doibase 10.1016/j.ppnp.2026.104234}
  {\bibfield  {journal} {\bibinfo  {journal} {Prog. Part. Nucl. Phys.}\
  }\textbf {\bibinfo {volume} {148}},\ \bibinfo {pages} {104234} (\bibinfo
  {year} {2026})},\ \Eprint {http://arxiv.org/abs/2503.00075} {arXiv:2503.00075
  [nucl-th]} \BibitemShut {NoStop}%
\bibitem [{\citenamefont {Hatsuda}\ and\ \citenamefont
  {Kunihiro}(1994)}]{Hatsuda:1994pi}%
  \BibitemOpen
  \bibfield  {author} {\bibinfo {author} {\bibfnamefont {T.}~\bibnamefont
  {Hatsuda}}\ and\ \bibinfo {author} {\bibfnamefont {T.}~\bibnamefont
  {Kunihiro}},\ }\href {\doibase 10.1016/0370-1573(94)90022-1} {\bibfield
  {journal} {\bibinfo  {journal} {Phys. Rept.}\ }\textbf {\bibinfo {volume}
  {247}},\ \bibinfo {pages} {221} (\bibinfo {year} {1994})},\ \Eprint
  {http://arxiv.org/abs/hep-ph/9401310} {arXiv:hep-ph/9401310} \BibitemShut
  {NoStop}%
\bibitem [{\citenamefont {Kapusta}\ and\ \citenamefont
  {Gale}(2011)}]{Kapusta:2006pm}%
  \BibitemOpen
  \bibfield  {author} {\bibinfo {author} {\bibfnamefont {J.~I.}\ \bibnamefont
  {Kapusta}}\ and\ \bibinfo {author} {\bibfnamefont {C.}~\bibnamefont {Gale}},\
  }\href {\doibase 10.1017/CBO9780511535130} {\emph {\bibinfo {title}
  {{Finite-temperature field theory: Principles and applications}}}},\
  Cambridge Monographs on Mathematical Physics\ (\bibinfo  {publisher}
  {Cambridge University Press},\ \bibinfo {year} {2011})\BibitemShut {NoStop}%
\bibitem [{\citenamefont {Bellac}(2011)}]{Bellac:2011kqa}%
  \BibitemOpen
  \bibfield  {author} {\bibinfo {author} {\bibfnamefont {M.~L.}\ \bibnamefont
  {Bellac}},\ }\href {\doibase 10.1017/CBO9780511721700} {\emph {\bibinfo
  {title} {{Thermal Field Theory}}}},\ Cambridge Monographs on Mathematical
  Physics\ (\bibinfo  {publisher} {Cambridge University Press},\ \bibinfo
  {year} {2011})\BibitemShut {NoStop}%
\bibitem [{\citenamefont {Cohen}(2003)}]{Cohen:2003kd}%
  \BibitemOpen
  \bibfield  {author} {\bibinfo {author} {\bibfnamefont {T.~D.~.}\ \bibnamefont
  {Cohen}},\ }\href {\doibase 10.1103/PhysRevLett.91.222001} {\bibfield
  {journal} {\bibinfo  {journal} {Phys. Rev. Lett.}\ }\textbf {\bibinfo
  {volume} {91}},\ \bibinfo {pages} {222001} (\bibinfo {year} {2003})},\
  \Eprint {http://arxiv.org/abs/hep-ph/0307089} {arXiv:hep-ph/0307089}
  \BibitemShut {NoStop}%
\bibitem [{\citenamefont {Aoki}\ \emph {et~al.}(2006)\citenamefont {Aoki},
  \citenamefont {Fodor}, \citenamefont {Katz},\ and\ \citenamefont
  {Szabo}}]{Aoki:2006we}%
  \BibitemOpen
  \bibfield  {author} {\bibinfo {author} {\bibfnamefont {Y.}~\bibnamefont
  {Aoki}}, \bibinfo {author} {\bibfnamefont {Z.}~\bibnamefont {Fodor}},
  \bibinfo {author} {\bibfnamefont {S.~D.}\ \bibnamefont {Katz}}, \ and\
  \bibinfo {author} {\bibfnamefont {K.~K.}\ \bibnamefont {Szabo}},\ }\href
  {\doibase 10.1016/j.physletb.2006.10.021} {\bibfield  {journal} {\bibinfo
  {journal} {Phys. Lett. B}\ }\textbf {\bibinfo {volume} {643}},\ \bibinfo
  {pages} {46} (\bibinfo {year} {2006})},\ \Eprint
  {http://arxiv.org/abs/hep-lat/0609068} {arXiv:hep-lat/0609068} \BibitemShut
  {NoStop}%
\bibitem [{\citenamefont {Bhattacharya}\ \emph {et~al.}(2014)\citenamefont
  {Bhattacharya} \emph {et~al.}}]{Bhattacharya:2014ara}%
  \BibitemOpen
  \bibfield  {author} {\bibinfo {author} {\bibfnamefont {T.}~\bibnamefont
  {Bhattacharya}} \emph {et~al.} (\bibinfo {collaboration} {HotQCD}),\ }\href
  {\doibase 10.1103/PhysRevLett.113.082001} {\bibfield  {journal} {\bibinfo
  {journal} {Phys. Rev. Lett.}\ }\textbf {\bibinfo {volume} {113}},\ \bibinfo
  {pages} {082001} (\bibinfo {year} {2014})},\ \Eprint
  {http://arxiv.org/abs/1402.5175} {arXiv:1402.5175 [hep-lat]} \BibitemShut
  {NoStop}%
\bibitem [{\citenamefont {Bazavov}\ \emph {et~al.}(2014)\citenamefont {Bazavov}
  \emph {et~al.}}]{HotQCD:2014kol}%
  \BibitemOpen
  \bibfield  {author} {\bibinfo {author} {\bibfnamefont {A.}~\bibnamefont
  {Bazavov}} \emph {et~al.} (\bibinfo {collaboration} {HotQCD}),\ }\href
  {\doibase 10.1103/PhysRevD.90.094503} {\bibfield  {journal} {\bibinfo
  {journal} {Phys. Rev. D}\ }\textbf {\bibinfo {volume} {90}},\ \bibinfo
  {pages} {094503} (\bibinfo {year} {2014})},\ \Eprint
  {http://arxiv.org/abs/1407.6387} {arXiv:1407.6387 [hep-lat]} \BibitemShut
  {NoStop}%
\bibitem [{\citenamefont {Bratovic}\ \emph {et~al.}(2013)\citenamefont
  {Bratovic}, \citenamefont {Hatsuda},\ and\ \citenamefont
  {Weise}}]{Bratovic:2012sd}%
  \BibitemOpen
  \bibfield  {author} {\bibinfo {author} {\bibfnamefont {N.~M.}\ \bibnamefont
  {Bratovic}}, \bibinfo {author} {\bibfnamefont {T.}~\bibnamefont {Hatsuda}}, \
  and\ \bibinfo {author} {\bibfnamefont {W.}~\bibnamefont {Weise}},\ }\href
  {\doibase 10.1016/j.physletb.2013.01.003} {\bibfield  {journal} {\bibinfo
  {journal} {Phys. Lett. B}\ }\textbf {\bibinfo {volume} {719}},\ \bibinfo
  {pages} {131} (\bibinfo {year} {2013})},\ \Eprint
  {http://arxiv.org/abs/1204.3788} {arXiv:1204.3788 [hep-ph]} \BibitemShut
  {NoStop}%
\bibitem [{\citenamefont {Steinheimer}\ and\ \citenamefont
  {Schramm}(2011)}]{Steinheimer:2010sp}%
  \BibitemOpen
  \bibfield  {author} {\bibinfo {author} {\bibfnamefont {J.}~\bibnamefont
  {Steinheimer}}\ and\ \bibinfo {author} {\bibfnamefont {S.}~\bibnamefont
  {Schramm}},\ }\href {\doibase 10.1016/j.physletb.2010.12.029} {\bibfield
  {journal} {\bibinfo  {journal} {Phys. Lett. B}\ }\textbf {\bibinfo {volume}
  {696}},\ \bibinfo {pages} {257} (\bibinfo {year} {2011})},\ \Eprint
  {http://arxiv.org/abs/1011.2462} {arXiv:1011.2462 [hep-ph]} \BibitemShut
  {NoStop}%
\bibitem [{\citenamefont {Masuda}\ \emph {et~al.}(2016)\citenamefont {Masuda},
  \citenamefont {Hatsuda},\ and\ \citenamefont {Takatsuka}}]{Masuda:2015jka}%
  \BibitemOpen
  \bibfield  {author} {\bibinfo {author} {\bibfnamefont {K.}~\bibnamefont
  {Masuda}}, \bibinfo {author} {\bibfnamefont {T.}~\bibnamefont {Hatsuda}}, \
  and\ \bibinfo {author} {\bibfnamefont {T.}~\bibnamefont {Takatsuka}},\ }\href
  {\doibase 10.1093/ptep/ptv170} {\bibfield  {journal} {\bibinfo  {journal}
  {Prog. Theor. Exp. Phys.}\ }\textbf {\bibinfo {volume} {2016}},\ \bibinfo
  {pages} {013D01} (\bibinfo {year} {2016})},\ \Eprint
  {http://arxiv.org/abs/1508.04861} {arXiv:1508.04861 [nucl-th]} \BibitemShut
  {NoStop}%
\end{thebibliography}
%

\end{document}